\PassOptionsToPackage{section}{placeins}
\PassOptionsToPackage{pdfversion=1.7}{hyperref}
\documentclass[10pt]{njulink}
\njulinksetup{
  short-title={TestPrism: Rethinking Test Evaluation Beyond a Single Reference}
}

\usepackage{amsmath,amsfonts,bm}

\def\eqref#1{equation~\ref{#1}}

\def\1{\bm{1}}

\DeclareMathAlphabet{\mathsfit}{\encodingdefault}{\sfdefault}{m}{sl}
\SetMathAlphabet{\mathsfit}{bold}{\encodingdefault}{\sfdefault}{bx}{n}

\newcommand{\benchTaskCount}{300}
\newcommand{\benchSourceCount}{17}
\newcommand{\benchPatchCount}{3000}

\usepackage{pifont}
\usepackage{wrapfig}
\usepackage{needspace}
\newcommand{\parabf}[1]{\noindent\textbf{#1}\enspace}

\definecolor{answerblue}{RGB}{243,248,252}
\definecolor{tableaccent}{RGB}{240,235,248}
\definecolor{rqanswerback}{RGB}{237,245,253}
\definecolor{rqanswerrule}{RGB}{57,119,180}
\definecolor{promptink}{RGB}{38,63,86}
\definecolor{promptback}{RGB}{247,249,252}
\definecolor{promptborder}{RGB}{185,203,220}
\newtcolorbox{promptcard}[2]{%
  enhanced,
  breakable,
  width=\linewidth,
  colback=promptback,
  colframe=promptborder,
  colbacktitle=promptink,
  coltitle=white,
  fonttitle=\small\sffamily\bfseries,
  title={\textcolor{promptborder}{#1}\enspace #2\hfill{\scriptsize\mdseries PROMPT EXCERPTS}},
  title after break={\textcolor{promptborder}{#1}\enspace #2\hfill{\scriptsize\mdseries CONTINUED}},
  fontupper=\small\ttfamily,
  boxrule=0.45pt,
  titlerule=0pt,
  arc=2.5pt,
  boxsep=0pt,
  left=10pt,right=10pt,top=8pt,bottom=8pt,
  toptitle=6pt,bottomtitle=6pt,
  pad at break*=2mm,
  before upper={%
    \renewenvironment{quote}{\par\noindent\begin{minipage}{\linewidth}}{\end{minipage}\par}%
    \raggedright}}
\newcommand{\promptdivider}{%
  \par%
  {\color{promptborder!65!white}\hrule height 0.35pt}%
  }
\newcommand{\promptsubheading}[1]{%
  \par{\footnotesize\sffamily\bfseries\textcolor{promptink}{#1}}\par\nobreak}
\newtcolorbox{rqanswer}[1]{%
  enhanced,
  width=\linewidth,
  colback=rqanswerback,
  frame hidden,
  boxrule=0pt,
  sharp corners,
  borderline west={2pt}{0pt}{rqanswerrule},
  boxsep=3pt,
  left=5pt,right=3pt,top=2pt,bottom=2pt,
  before upper={\textbf{Answer to RQ#1:}\enspace}}

\newcolumntype{C}{>{\centering\arraybackslash}X}
\usepackage{flafter}

\title{TestPrism: Rethinking Test Evaluation Beyond a Single Reference}
\author[1,*]{Han Li}
\author[1,*]{Lingxiang Hu}
\author[1,*]{Jiacheng Huang}
\author[1]{Ziqian Jiang}
\author[1]{Jingkai Luo}
\author[1]{Wei Gao}
\author[2]{Yunfan Tan}
\author[3]{Zun Wang}
\author[1,\dagger]{Jiaheng Liu}
\affiliation[1]{Nanjing University}
\affiliation[2]{Hong Kong Polytechnic University}
\affiliation[3]{Tencent}
\contribution[*]{Equal contribution.}
\contribution[\dagger]{Corresponding author.}
\date{\today}

\abstract{Large language model (LLM) coding agents have advanced test generation across
diverse programming tasks. However, the common practice of evaluating tests
against a single reference solution overlooks alternative valid
implementations and can overstate test quality. We introduce \textbf{TestPrism},
comprising \benchTaskCount{} test tasks from \benchSourceCount{} sources and
\benchPatchCount{} candidate implementations, evenly split between valid and
invalid solutions. Its primary metric, \textbf{Joint Success Function},
requires the generated tests to fail on the initial program state,
accept every valid candidate, and reject every invalid candidate.
Across fourteen baseline coding agent configurations,
Joint Success Function reaches only 28.00\%, whereas single reference success reaches 59.67\%.
Our analysis reveals missed behaviors, unsupported assertions, and faulty
test construction. To address these weaknesses, we introduce
\textbf{TestHelix}, which combines heterogeneous synthesis of test and
repair pairs with peer cross validation and recursive self improvement
(RSI).
Across two models, TestHelix improves Joint Success Function by 8.67 to 9.00 percentage points
over the native harness comparators in the TestHelix evaluation.
}

\begin{document}
\maketitle

\begingroup
\renewcommand{\parabf}[1]{\noindent\textbf{#1}\enspace}

\section{Introduction}

Large language model (LLM) coding agents, such as Codex and Claude Code, can
navigate repositories, modify implementations, and validate changes through
execution feedback~\citep{yang2024sweagent,wang2024codeact,wang2025openhands,pan2026prometheus,li2026solvita}.
These capabilities also support automated test generation and
improvement~\citep{chen2024chatunitest,alshahwan2024meta}. Reliable tests must
generalize across valid implementations while remaining sensitive to
specification violations~\citep{zhu1997adequacy,barr2015oracle}.

\providecommand{\cmark}{\textcolor{green!50!black}{\ensuremath{\checkmark}}}
\providecommand{\xmark}{\textcolor{red!80!black}{\ensuremath{\times}}}
\providecommand{\relatedmark}{\textcolor{orange!70!black}{\ensuremath{\dagger}}}

\begin{table}[t]
\centering
\caption{\textbf{Scope and evaluation targets of test generation benchmarks.}}
\label{tab:compare_benches}
\begingroup
\footnotesize
\definecolor{comparisonheader}{RGB}{40,80,140}
\definecolor{comparisonlocal}{RGB}{245,240,255}
\definecolor{comparisonrepo}{RGB}{240,255,240}
\definecolor{comparisonstripe}{RGB}{248,248,252}
\definecolor{comparisonours}{RGB}{220,190,255}
\renewcommand{\cmark}{\textcolor{green!50!black}{\ding{51}}}
\renewcommand{\xmark}{\textcolor{red}{\ding{55}}}
\newtcbox{\scopefile}{on line,arc=3pt,colback=yellow!30,colframe=yellow!90,
  boxsep=0pt,left=4pt,right=4pt,top=2pt,bottom=2pt,boxrule=0.8pt,
  fontupper=\scriptsize\bfseries,nobeforeafter}
\newtcbox{\scopemultifile}{on line,arc=3pt,colback=green!10,colframe=green!60!black,
  boxsep=0pt,left=4pt,right=4pt,top=2pt,bottom=2pt,boxrule=0.8pt,
  fontupper=\scriptsize\bfseries\color{green!50!black},nobeforeafter}
\newtcbox{\scoperepo}{on line,arc=3pt,colback=red!8,colframe=red!60,
  boxsep=0pt,left=4pt,right=4pt,top=2pt,bottom=2pt,boxrule=0.8pt,
  fontupper=\scriptsize\bfseries\color{red!70!black},nobeforeafter}
\setlength{\tabcolsep}{3pt}
\resizebox{\linewidth}{!}{%
\begin{tabular}{@{}>{\columncolor{white}[0pt][\tabcolsep]}lrrccccccc>{\columncolor{white}[\tabcolsep][0pt]}c@{}}
\toprule
\textbf{Benchmark} & \textbf{Tasks} & \textbf{Variants} & \textbf{Languages}
& \textbf{RAR} & \textbf{VAR} & \textbf{IRR} & \textbf{JSF}
& $\boldsymbol{\Delta\mathcal{C}}$ & $\mathbf{EC}$ & \textbf{Scope} \\
\midrule
\multicolumn{11}{c}{\cellcolor{comparisonlocal}\textit{\textcolor{comparisonheader}{\textbf{Block, File, and Multifile Benchmarks}}}} \\
\midrule
TestGenEval & 1,210 & \cmark & 1 & \xmark & \xmark & \relatedmark & \xmark & \xmark & \xmark & \scopefile{File} \\
\rowcolor{comparisonstripe}
CLOVER & 845 & \xmark & 1 & \xmark & \xmark & \xmark & \xmark & \xmark & \xmark & \scopefile{Block/file} \\
MultiFileTest & 60 & \xmark & 3 & \xmark & \xmark & \xmark & \xmark & \xmark & \xmark & \scopemultifile{Multifile} \\
\midrule
\multicolumn{11}{c}{\cellcolor{comparisonrepo}\textit{\textcolor{comparisonheader}{\textbf{Repository Benchmarks}}}} \\
\midrule
SWT Bench & 1,983 & \xmark & 1 & \cmark & \xmark & \xmark & \xmark & \relatedmark & \xmark & \scoperepo{Repo.} \\
\rowcolor{comparisonstripe}
TDD Bench Verified & 449 & \xmark & 1 & \cmark & \xmark & \xmark & \xmark & \relatedmark & \xmark & \scoperepo{Repo.} \\
OmniCode & 396 & 1,109 & 3 & \xmark & \xmark & \relatedmark & \xmark & \xmark & \xmark & \scoperepo{Repo.} \\
\rowcolor{comparisonstripe}
SWE Mutation & 800 & 2,636 & 10 & \cmark & \xmark & \relatedmark & \xmark & \xmark & \xmark & \scoperepo{Repo.} \\
\midrule
\rowcolor{comparisonours}
\textbf{TestPrism (Ours)} & \textbf{\benchTaskCount{}} & \textbf{2,700} & \textbf{6}
& \cmark & \cmark & \cmark & \cmark & \cmark & \cmark
& \scopefile{Block/file}\;\scopemultifile{Multifile}\;\scoperepo{Repo.} \\
\bottomrule
\end{tabular}%
}
\endgroup
\end{table}

\paragraph{Limitations of Existing Benchmarks.}
Test generation benchmarks span function level tasks, as in TestEval
\citep{wang2024testeval}, and repository issues, as in SWT Bench
\citep{mundler2024swtbench}. A common success criterion remains
\textbf{single reference evaluation}: a suite must fail on the initial program
and pass on one gold solution~\citep{mundler2024swtbench,ahmed2024tddbench}.
Yet a requirement can admit multiple valid
implementations and plausible incorrect ones. A test can pass this check
while missing required behavior~\citep{qi2015patchplausibility,liu2023evalplus}
or imposing unnecessary implementation constraints.
We therefore ask: \textit{How robust is the evaluation of generated
tests to variation across implementations?}
Table~\ref{tab:compare_benches} compares existing benchmarks; counting
conventions and target mappings appear in
Appendix~\ref{sec:benchmark-comparison-details}.

\paragraph{TestPrism: Beyond Single Reference Evaluation.}
TestPrism contains \benchTaskCount{} test tasks from \benchSourceCount{}
sources across six programming languages, covering repository maintenance,
algorithmic programming, and performance optimization. We adapt executable
tasks to a shared interface,
enrich their candidate patch sets, verify and select diverse implementations,
and convert the tasks for test generation. For each test task, we retain the
reference and select four valid alternatives and five invalid implementations
through explicit diversity checks.
Building on this pipeline, we construct TestPrism with 2,000 tasks from
\benchSourceCount{} source benchmarks, partitioned into 1,700 training tasks
and \benchTaskCount{} test tasks.
The test set contains \benchPatchCount{} candidate implementations and spans
six task categories and 34 subcategories across 67 repositories and projects.
This task diversity supports assessing how well
coding agents generalize test generation across programming scenarios.
The disjoint training set supports adapting test generation strategies in
TestHelix's recursive self improvement (RSI) stage.

Our evaluation protocol executes each
submitted suite on the initial state and every candidate. Its primary metric,
\textbf{Joint Success Function}, requires initial failure and agreement
with every candidate label. Restricting the panel to the reference recovers
single reference evaluation as a minimal approximation. Complementary
acceptance, rejection, and coverage metrics locate behavioral disagreements
and the code exercised by the tests.

\paragraph{TestHelix.}
We introduce TestHelix to address these weaknesses through heterogeneous
synthesis of test and repair pairs. Its inner workflow establishes
consistency within each pair, then uses peer cross validation, comprising
cross execution, semantic review, and corrective feedback, to check and correct tests
across pairs. Its outer RSI loop proposes test generation strategies on
training tasks and validates one frozen challenger on a disjoint set before
benchmark evaluation.

\paragraph{Takeaways.}
We evaluate fourteen coding agent configurations spanning six model families.
Our analysis reveals four key findings:
\ding{182}~\textbf{Reliable test generation remains far from solved by
frontier models.} The strongest baseline reaches only 28.00\% Joint Success Function, compared
with 59.67\% single reference success (Table~\ref{tab:benchmark300-main}).
Coverage tracks success: change and entry coverage correlate with Joint Success Function across
configurations ($r=0.91$ and $0.96$).
\ding{183}~\textbf{Joint Success Function is a dynamic measure of reliability across
implementations.} Larger candidate panels impose additional behavioral
requirements, as shown in Figure~\ref{fig:rq2-patch-count}.
Single reference evaluation is its weakest special case and an upper bound
on expanded panel success.
\ding{184}~\textbf{Failures span coverage gaps, faulty checks, and execution
logic errors.} We sample 300 failed trajectories, 50 from the
highest Joint Success Function configuration in each of six model families, and analyze 286
after excluding 14 timeout trajectories.
The category shares among the analyzed trajectories are 34.6\%, 50.0\%, and
15.4\%, respectively (Figure~\ref{fig:error-taxonomy}), explaining limited
reliability through omissions, misjudgments, and execution failures.
Missing inputs and weak assertions can pass single reference evaluation
while accepting invalid patches.
\ding{185}~\textbf{Heterogeneous synthesis helps tests generalize across
implementations.} With peer cross validation using diverse repairs to check and correct tests
and RSI adapting test generation strategies across tasks, TestHelix raises Joint Success Function
over Single by 11.67 to 12.00 percentage points under matched inference budget
caps (Table~\ref{tab:testhelix-rsi}).

\paragraph{Contributions.}
Our work makes three contributions:
\begin{itemize}
\item We construct \textbf{TestPrism}, with 1,700 training tasks and
\benchTaskCount{} test tasks from \benchSourceCount{} sources, providing
\benchPatchCount{} labeled implementations for evaluation. Its Joint Success Function criterion
extends single reference evaluation to assess whether generated tests accept
valid alternatives and reject incorrect implementations.
\item We evaluate fourteen coding agent configurations across six model
families and develop an \textbf{error taxonomy} from 286 retained failed
trajectories spanning their highest Joint Success Function configurations. These
complementary analyses characterize current agent
performance and identify specific mistakes that cause generated tests to miss
invalid behavior or reject valid alternatives, including failures hidden by
single reference evaluation.
\item We introduce \textbf{TestHelix}, which combines heterogeneous synthesis
with peer cross validation and RSI. It checks and corrects tests through cross execution and semantic review,
and adapts reusable test generation strategies to improve reliability across
implementations under matched inference budget caps.
\end{itemize}

\section{TestPrism}
\label{sec:testprism}

To evaluate whether generated tests generalize beyond a single reference,
we construct TestPrism through four steps: \textbf{(i) task screening and
adaptation}, \textbf{(ii) task selection}, \textbf{(iii) patch generation and
validation}, and \textbf{(iv) diversity filtering and task conversion}.
Each test instance pairs a test generation task with a curated panel of valid
and invalid implementations. Figure~\ref{fig:construction} outlines the pipeline.

\begin{figure}[htbp]
\centering
\includegraphics[pagebox=mediabox,width=0.90\linewidth]{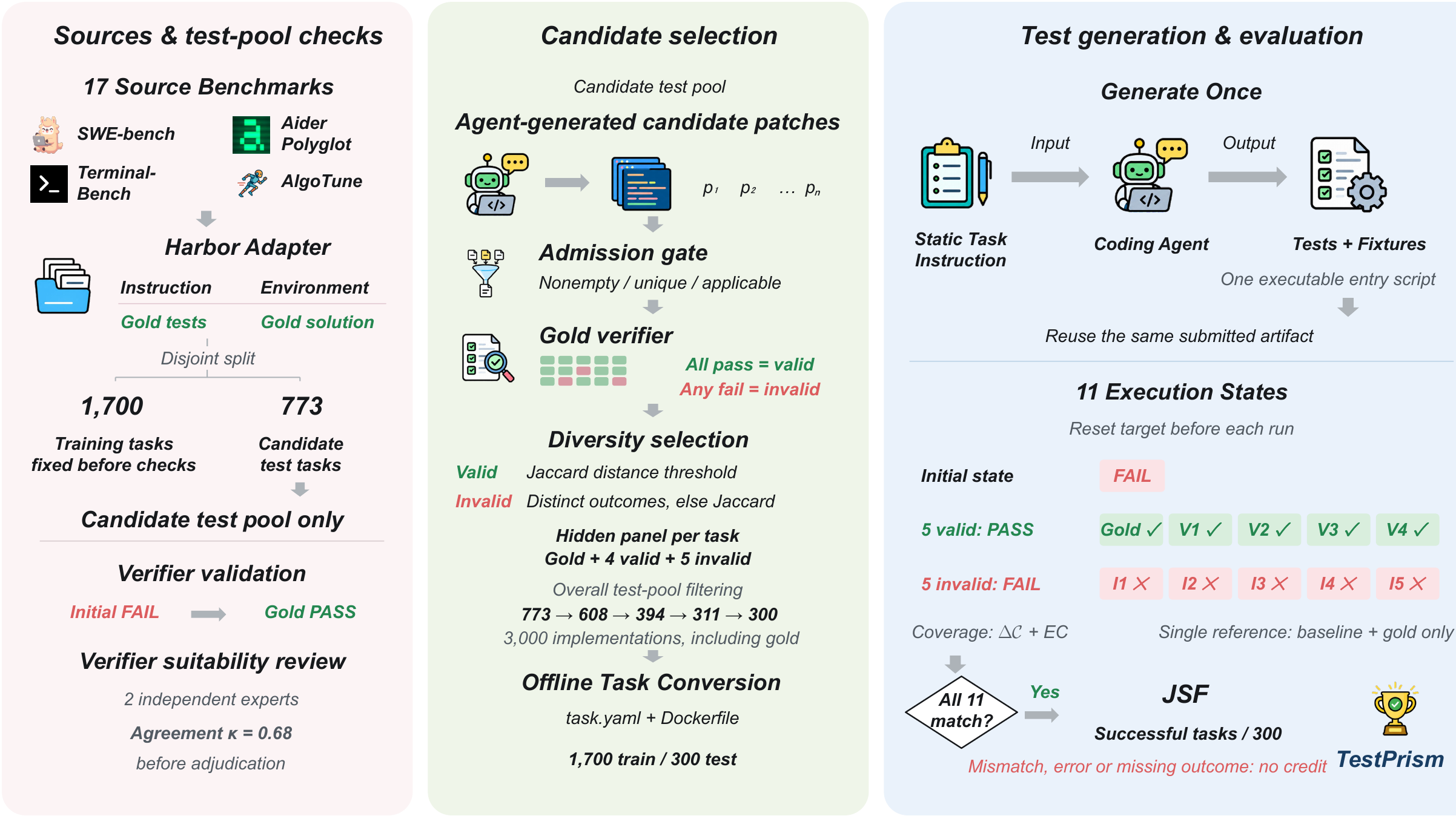}
\caption{\textbf{Overview of TestPrism construction and evaluation.}}
\label{fig:construction}
\end{figure}
\subsection{Step 1: Task Screening and Adaptation}
\label{sec:adapters}

We curate tasks from \benchSourceCount{} public benchmarks, including the
widely used SWE bench and Terminal Bench
families~\citep{jimenez2024swebench,merrill2026terminalbench}, requiring traceable test
provenance and documented validation (Appendix~\ref{app:source-test-validation}).
We retain coding tasks whose requirements can be checked with executable tests,
excluding unusable Docker images and irreproducible environments. Adapters
map tasks to Harbor's instruction, environment, tests, and solution components
while preserving requirements, initial states, verifiers, reference solutions,
dependencies, services, and execution limits.

\subsection{Step 2: Task Selection}
\label{sec:selection}

We screen tasks using \textbf{Solution Size}, the effective lines of
\texttt{solution.sh}; \textbf{Test Count}, the enumerated test functions or
metadata checks; and \textbf{Test Density}, this count per file changed by
the reference solution.
Lower bounds are calibrated to each source's own distribution, and source
quotas balance representation across the disjoint training and candidate test
sets. Appendix~\ref{app:task-selection} reports the source-specific
screening thresholds. We retain 1,700 training
tasks for TestHelix's recursive self improvement (RSI); only candidate test
tasks undergo the remaining patch generation and filtering steps.

\subsection{Step 3: Patch Generation and Validation}
\label{sec:panels}

\parabf{Verifier validation.}
All source benchmarks provide test suites validated through execution checks,
expert review, or official competition evaluation. Our additional validation
round checks each adapted verifier before candidate generation, requiring
failure on the initial state and success on the reference solution. Two
experts use development experience to independently assess task requirements
and source verifiers for potential overfitting and coverage gaps. Discussion
clarifies disagreements; tasks with unresolved concerns are excluded.
Independent verifier usability judgments achieve Cohen's $\kappa=0.68$.
Appendix~\ref{app:source-test-validation} documents the prior validation
evidence and our additional review.

\parabf{Patch generation.}
We generate candidate patches with different coding agents and models of
varying capabilities to capture alternative solutions and errors arising
during task solving, complementing mutation based fault
models~\citep{just2014mutants,jia2011mutation}.
We remove empty submissions, exact duplicates, and inapplicable
patches, then execute the source verifier in clean environments and record
outcomes for individual checks. A candidate is \textbf{valid} if every check
passes and \textbf{invalid} if any check fails. These labels remain fixed
when evaluating generated tests.

\subsection{Step 4: Diversity Filtering and Task Conversion}
\label{sec:testgen-adapter}

We assemble panels of one reference, four valid alternatives, and five
invalid implementations. The valid group must meet the mean AST distance
criterion. For invalid implementations, different sets of failed verifier
checks pass the diversity check directly; matching sets require the AST
Jaccard check. We refine panels through generation and replacement until
both group criteria hold, then apply source quotas in final sampling.

The \textbf{test generation adapter} converts each programming task into a
test authoring contract: submit tests, fixtures, and one executable entry
script in a designated artifact directory. The original requirement defines
the behavior to test; the candidate panel remains hidden during test
generation. Appendix~\ref{app:diversity-conversion} gives the filtering
stages and implementation details.

\subsection{TestPrism Statistics}
\label{sec:dataset-partitioning}

TestPrism contains 2,000 tasks collected and adapted from
\benchSourceCount{} source benchmarks, with 1,700 tasks in the training set
and \benchTaskCount{} in the test set. The partitions are disjoint.
The test set contains \benchPatchCount{} labeled implementations across six
programming languages and 67 repositories and projects.

The sources span repository maintenance, evolved function specifications,
olympiad programming, and numerical
optimization~\citep{xia2024evoeval,shi2024olympiad,press2025algotune}.
Appendix~\ref{app:sources} provides source and language statistics,
the source test count distribution (Figure~\ref{fig:test-count-distribution}),
the six task categories and 34 subcategories, and counting conventions.
Appendix~\ref{app:task-selection} details partitioning procedures and
eligibility criteria. The filtering
funnel appears in Appendix~\ref{app:task-selection}.
\subsection{Evaluation Protocol}
\label{sec:evaluation-protocol}

We run each submitted entry script on the initial state and every panel
implementation, resetting the target each time and recording one verdict.
For task $i$, $b_i$ is the initial state and $g_i$ the reference;
$\mathcal{V}_i$ contains $g_i$ and four valid alternatives, and
$\mathcal{I}_i$ contains five invalid candidates. Indicators $a_i(p)$ and
$r_i(p)$ denote recorded pass and fail outcomes on $p$; both are zero for
missing outcomes or recognized execution errors.

\parabf{Behavioral metrics.}
\textbf{Reference Acceptance Rate} measures single reference fail to
pass success over the fixed $N=\benchTaskCount{}$ tasks:
\[
\text{Reference Acceptance Rate}=\frac{1}{N}\sum_{i=1}^{N}r_i(b_i)\,a_i(g_i).
\]
\textbf{Valid Acceptance Rate} measures acceptance of valid alternatives;
\textbf{Invalid Rejection Rate} measures rejection of invalid candidates.
For each task, $A_i$ is the fraction of recorded valid alternatives
accepted, excluding $g_i$, and $R_i$ is the fraction of recorded invalid
candidates rejected. Let $\mathcal{T}_V$ and $\mathcal{T}_I$ contain the
tasks with defined $A_i$ and $R_i$, respectively. We average each task's
rate with equal weight:
\[
\text{Valid Acceptance Rate}=\frac{1}{\lvert\mathcal{T}_V\rvert}
  \sum_{i\in\mathcal{T}_V}A_i,
\qquad
\text{Invalid Rejection Rate}=\frac{1}{\lvert\mathcal{T}_I\rvert}
  \sum_{i\in\mathcal{T}_I}R_i.
\]
Neither rate is gated by Reference Acceptance Rate. Missing records are excluded; recorded errors
remain in the task denominators. An empty task denominator leaves its rate
undefined. All full-benchmark configurations have complete observations, so both
means cover all $N$ tasks
(Appendix~\ref{sec:behavioral-details}).
We also report their harmonic mean $H$, computed from both rates at the
precision shown in the tables.

\parabf{Coverage.}
\label{sec:coverage}
Following SWT Bench, \textbf{Change Coverage ($\Delta\mathcal{C}$)} measures
the fraction of executable reference changes exercised by generated tests;
trusted tests identify targets in the corresponding initial or reference
state~\citep{mundler2024swtbench}.
Following TestExplora, \textbf{Entry Coverage} measures coverage
of executable code within production entry spans identified by trusted tests
in the reference state, potentially beyond changed
lines~\citep{liu2026testexplora}.
Failed collection and empty denominators yield no numeric value; we report
collection counts (Appendix~\ref{sec:coverage-details}).

\parabf{Joint Success Function.}
\label{sec:jsf}
Our primary metric, \textbf{Joint Success Function}, takes candidate
patch sets $\mathcal{P}=(P_i)_{i=1}^{N}$, where
$g_i\in P_i\subseteq\mathcal{V}_i\cup\mathcal{I}_i$.
It requires initial failure, acceptance of every selected valid implementation,
and rejection of every selected invalid implementation:
\[
\text{Joint Success Function}(\mathcal{P})
=\frac{1}{N}\sum_{i=1}^{N}
  r_i(b_i)\,
  \prod_{p\in P_i\cap\mathcal{V}_i} a_i(p)\,
  \prod_{p\in P_i\cap\mathcal{I}_i} r_i(p).
\]
Reference sets $\mathcal{P}_{\mathrm{ref}}=(\{g_i\})_{i=1}^{N}$ recover Reference Acceptance Rate.
Unless otherwise specified, reported Joint Success Function values use the full panels,
$P_i=\mathcal{V}_i\cup\mathcal{I}_i$, fixed across systems.
\label{sec:populations}
Joint Success Function cannot exceed Reference Acceptance Rate; their gap
measures the fraction of tasks
that pass reference evaluation but fail joint evaluation, including missing
or erroneous outcomes.

\section{TestHelix}
\label{sec:testhelix}

To improve test reliability beyond a single reference implementation,
we propose TestHelix with three components
(Figure~\ref{fig:testhelix-architecture}).
\textbf{Heterogeneous synthesis} generates diverse test and repair pairs
to explore complementary behaviors and implementation hypotheses.
\textbf{Peer cross validation} checks tests against peer repairs to expose
and correct errors that agreement with one repair can miss.
\textbf{Recursive self improvement (RSI)} learns from execution feedback
across tasks to improve the test generation strategy.

\begin{figure}[htbp]
    \centering
    \includegraphics[width=\linewidth]{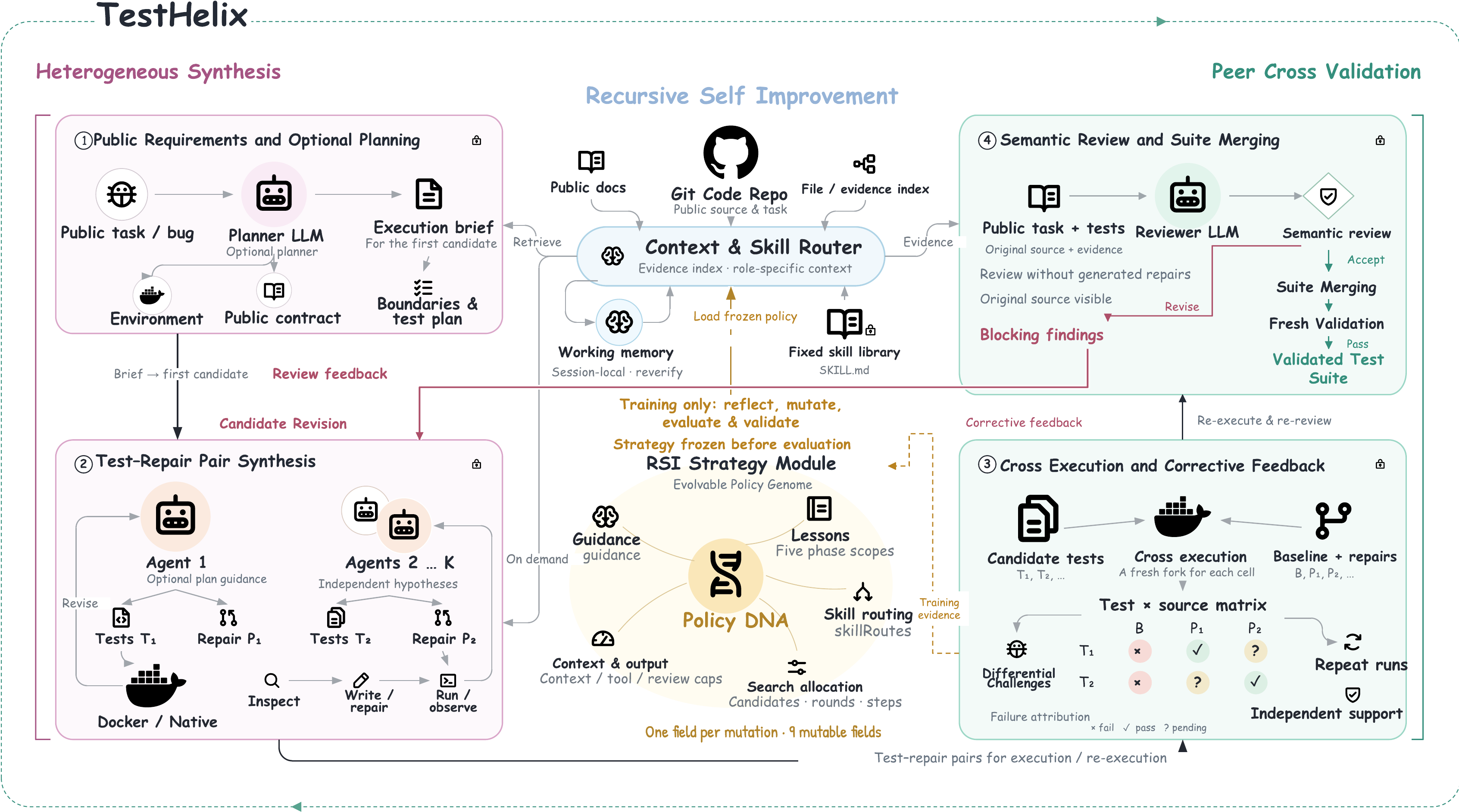}
    \caption{\textbf{TestHelix architecture.}}
    \label{fig:testhelix-architecture}
\end{figure}

\subsection{Heterogeneous Synthesis}

\textbf{Diversity.} We encourage diversity by assigning distinct exploration strategies to
isolated coding agents. Each agent independently develops a test patch $T_k$
and a repair patch $P_k$ from public task requirements.
\textbf{Internal consistency.} We enforce internal consistency within each
pair by requiring $T_k$ to reproducibly fail on the
initial program and pass on $P_k$. Agents use local execution feedback to
refine their pairs until this condition holds or the budget is exhausted.
Appendix~\ref{app:testhelix-hs} details the synthesis prompts and branch
strategies.

\subsection{Peer Cross Validation}

\textbf{Cross execution.} We execute $T_i$ against internally consistent
peer repairs $P_j$ to expose overly restrictive tests or incomplete repairs.
\textbf{Semantic review.} An independent reviewer checks assertions against
public requirements without seeing generated repairs and supports every
blocking finding with public evidence.
\textbf{Corrective feedback.} We use reviewed counterexamples to correct
the pairs. When distinct qualified repairs pass the approved suites, we
use differential challenges to find inputs that separate
them~\citep{yang2011csmith}. We repeat
execution and review after revision.
\textbf{Suite merging.} We combine compatible suites with common passing
witnesses and validate the merged suite in a fresh environment.
Appendix~\ref{app:testhelix-pcv} documents the review prompt and evidence
boundary.

\subsection{Recursive Self Improvement}
\label{sec:testhelix-rsi}

We initialize TestHelix with a basic strategy. However, improving test
generation performance across heterogeneous tasks involves a large semantic
search space. We use RSI to refine
the strategy within the existing workflow while keeping the pair count and
inference budget fixed. We optimize Reference Acceptance Rate for a denser
success signal, since the stricter Joint Success Function
can leave incremental harness improvements unrewarded. Heterogeneous
synthesis and peer cross validation target the broader discrimination
across implementations required by Joint Success Function.

We use AutoSaddler~\citep{park2026autosaddler} to optimize the harness
strategy through reflective evolution. It selects a parent strategy,
diagnoses its execution traces, and proposes a revision. The parent and
child are evaluated on the same tasks using Reference Acceptance Rate; their performance guides
selection, while reflection extracts reusable lessons for later rounds.
Appendix~\ref{app:testhelix-development} details the optimization setup.

\section{Can Coding Agents Generate Reliable Tests Across Implementations?}
\label{sec:experiments}

To assess test reliability across implementations, we investigate the following
Research Questions (RQs): \textbf{RQ1} benchmarks coding agents and relates
coverage to success; \textbf{RQ2} examines how expanding diverse candidate
patch sets affects Joint Success Function; \textbf{RQ3} characterizes test generation failures; and \textbf{RQ4}
evaluates TestHelix and its components. Evaluations follow the shared protocol
in Section~\ref{sec:evaluation-protocol}.

\subsection{RQ1: Benchmarking Coding Agents}
\label{sec:rq1-performance}

\textit{How reliably do current coding agents generate tests that distinguish
valid and invalid implementations?}

Table~\ref{tab:benchmark300-main} benchmarks coding agents on all
\benchTaskCount{} tasks, reporting the highest Joint Success Function configuration from each
of six model families among fourteen evaluated configurations.
The main comparison uses native coding agents to reflect the close coupling
between model training and harness design. Appendix~\ref{sec:baseline-configurations}
provides the remaining eight configurations, including earlier versions and
model variants. We also examine how coverage and the balance between
acceptance and rejection relate to Joint Success Function across all fourteen configurations;
Appendix~\ref{sec:coverage-analysis-details} provides the correlation analysis.

\begin{table}[!htbp]
\centering
\caption{\textbf{Performance of the highest Joint Success Function configuration from each model family.}}
\label{tab:benchmark300-main}
\begingroup
\footnotesize
\setlength{\tabcolsep}{3pt}
\begin{tabular*}{\linewidth}{@{\extracolsep{\fill}}lrrrrrrr}
\toprule[1pt]
\textbf{Model}
& \multicolumn{1}{c}{\textbf{Reference}}
& \multicolumn{3}{c}{\textbf{Candidates}}
& \multicolumn{2}{c}{\textbf{Coverage}}
& \multicolumn{1}{c}{\textbf{Joint}} \\
\cmidrule(lr){2-2}\cmidrule(lr){3-5}\cmidrule(lr){6-7}\cmidrule(l){8-8}
& RAR & VAR & IRR & $H$ & $\Delta\mathcal{C}$ & $\mathrm{EC}$ & \textbf{JSF} \\
\midrule
Claude Fable 5.1 & \textbf{59.67\%} & 72.67\% & 79.33\% & \textbf{75.85\%} & 82.33\% & \textbf{54.33\%} & \cellcolor{tableaccent}\textbf{28.00\%} \\
Kimi K3 & 56.67\% & 63.67\% & 83.00\% & 72.06\% & \textbf{84.33\%} & 52.00\% & \cellcolor{tableaccent}24.67\% \\
Qwen3.8 Max & 54.67\% & \textbf{77.00\%} & 66.00\% & 71.08\% & 80.33\% & 50.33\% & \cellcolor{tableaccent}23.67\% \\
GPT 5.6 sol -max & 54.00\% & 61.33\% & 86.00\% & 71.60\% & 70.33\% & 49.00\% & \cellcolor{tableaccent}23.33\% \\
DeepSeek V4 Pro & 41.00\% & 57.67\% & 70.33\% & 63.37\% & 67.33\% & 43.33\% & \cellcolor{tableaccent}13.33\% \\
GLM 5.3 & 43.00\% & 50.33\% & \textbf{86.33\%} & 63.59\% & 64.67\% & 44.33\% & \cellcolor{tableaccent}11.33\% \\
\bottomrule[1pt]
\end{tabular*}
\endgroup
\end{table}

Reference success often fails to extend to other implementations. Claude
Fable 5.1 leads Joint Success Function at 28.00\%, compared with 59.67\%
against the reference. Five of the six configurations score higher on
Invalid Rejection Rate than Valid Acceptance Rate, indicating difficulty
accepting permissible implementations. GLM 5.3 has the highest rejection
rate but the lowest acceptance and joint success rates: defect rejection
alone is insufficient. Across all fourteen configurations, the positive
association between $H$ and Joint Success Function supports balancing both
requirements. Change Coverage and Entry Coverage also correlate with joint
success (Pearson $r=0.91$ and $0.96$). Reaching relevant behavior helps, but
does not ensure that assertions judge it
correctly~\citep{inozemtseva2014coverage,zhu1997adequacy}.

\begin{rqanswer}{1}
\textbf{Reliable test generation remains far from solved by frontier models.}
The highest baseline Joint Success Function is only 28.00\%. Valid Acceptance Rate and Invalid Rejection Rate expose complementary
weaknesses in accepting valid alternatives and rejecting defective patches;
their harmonic mean $H$ and coverage both track joint success.
\end{rqanswer}

\Needspace{6\baselineskip}
\subsection{RQ2: Patch Diversity and Joint Success}
\label{sec:rq2-reference-gap}

\textit{How does expanding a diverse candidate patch set affect Joint Success Function?}

Figure~\ref{fig:rq2-patch-count} shows Joint Success Function and candidate diversity
under three patch addition schedules.
We summarize diversity using mean pairwise Jaccard distance between AST feature sets.
For this diagnostic, we randomly sample 50 benchmark tasks and continue
the candidate construction pipeline to expand each task's standard ten
implementations to 15: the reference, seven additional valid alternatives,
and seven invalid candidates. We keep each submitted suite fixed and grow
its candidate set from the reference repair by adding valid alternatives,
invalid candidates, or the two in alternation. The alternating schedule starts
with an invalid addition and contains eight valid and seven invalid implementations
at $k=15$. The diagnostic equally weights the six model configurations in
Table~\ref{tab:benchmark300-main}, each evaluated on the same 50 tasks.
Candidate count and diversity
change together in this diagnostic; Appendix~\ref{sec:patch-count-diagnostic}
provides the construction and calculation details.

The results show two ways in which an expanded panel can remove
reference successes: tests may reject valid alternatives or accept invalid
candidates. Mean Joint Success Function starts at 49.33\%, reaches 38.10\% after five
valid additions and 30.49\% after five invalid additions, and reaches 24.33\%
after nine alternating additions. The complete pool endpoints are 37.33\%,
29.33\%, and 23.33\% for the valid, invalid, and alternating schedules.
The tests remain fixed while
the panel imposes additional acceptance and rejection requirements.

\noindent\begin{minipage}{\linewidth}
\centering
\includegraphics[width=\linewidth]{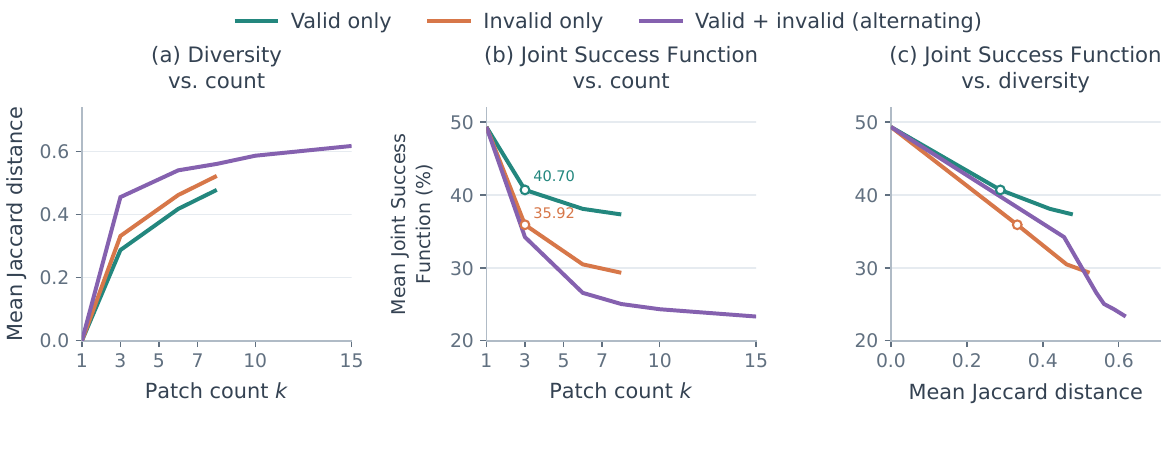}
\captionof{figure}{Joint Success Function and candidate diversity under three patch addition schedules.}
\label{fig:rq2-patch-count}
\end{minipage}

\begin{rqanswer}{2}
\textbf{Diverse patch additions can expose test errors left unchecked by
smaller panels.} Joint Success Function depends on which implementations a panel contains as
well as their number. Expanding the panel checks acceptance of valid
alternatives and rejection of defective candidates. Using only the reference
reduces the criterion to Reference Acceptance Rate.
\end{rqanswer}

\subsection{RQ3: Analysis of Test Generation Failures}
\label{sec:rq3-errors}

\textit{What error mechanisms occur in generated tests, and which remain hidden
under single reference evaluation?}

We sample 50 failed trajectories, each with a zero task-level joint success
indicator, from each configuration in Table~\ref{tab:benchmark300-main}.
After excluding 14 timeouts, two experts
independently annotate all 286 retained trajectories using seven error
subtypes (Cohen's $\kappa=0.81$).
Figure~\ref{fig:error-taxonomy} pairs the full benchmark metrics from
Table~\ref{tab:benchmark300-main} with failure mechanisms from this sample.
The radar axes are independently scaled: the displayed lower and upper
bounds define the center and outer polygon for each metric.
Appendix~\ref{sec:error-taxonomy-details} details expert annotation,
disagreement resolution, and subtype definitions.

\begin{figure}[!htbp]
\centering
\begin{minipage}[t]{0.40\linewidth}
\centering
\includegraphics[height=1.65in]{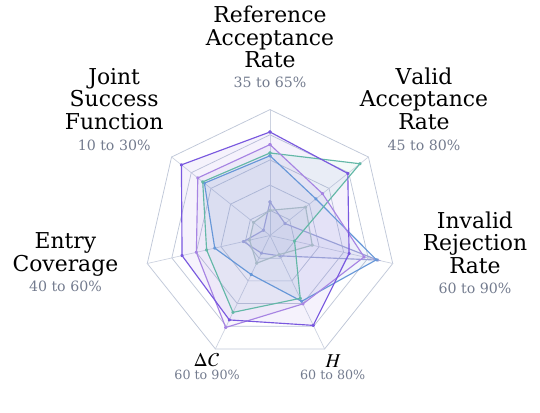}
\par
{\scriptsize (a) Model profiles\par}
\end{minipage}\hfill
\begin{minipage}[t]{0.58\linewidth}
\centering
\includegraphics[height=1.60in,trim=258.75bp 0bp 0bp 0bp,clip]{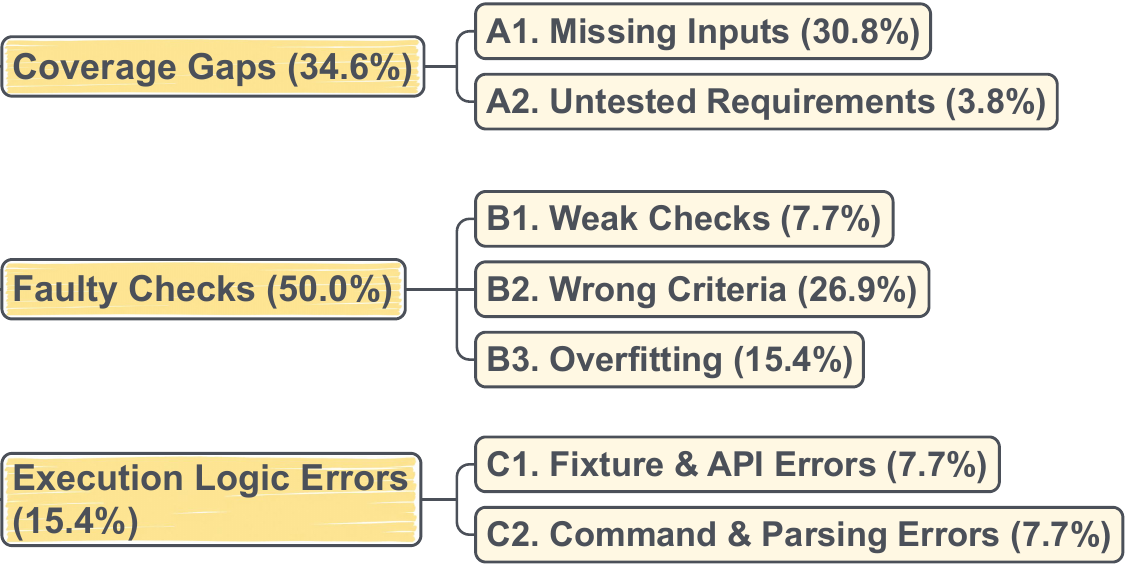}
\par
{\scriptsize (b) Test failure mechanisms\par}
\end{minipage}
\par
\includegraphics[width=\linewidth]{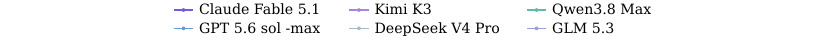}
\caption{Model profiles across all 300 benchmark tasks (a) and failure mechanisms
in 286 sampled failed trajectories after excluding 14 timeouts (b).}
\label{fig:error-taxonomy}
\end{figure}

The profiles in Figure~\ref{fig:error-taxonomy}(a) show contrasting strengths:
\mbox{GPT 5.6 sol -max} and GLM 5.3 reject invalid candidates more reliably
than they accept valid alternatives, whereas Qwen3.8 Max shows the reverse.
Panel (b) summarizes 99 coverage gaps (34.6\%), 143 faulty checks (50.0\%),
and 44 execution logic errors (15.4\%).
\textbf{Coverage gaps.} Tests can exercise an operation while missing the
input combinations that expose a defect. A poker suite never compares an
ace low straight against three of a kind, allowing a candidate that
misclassifies the straight as a high card to pass.
\textbf{Faulty checks.} The largest group shows that exercising relevant
behavior does not ensure that tests judge it correctly, consistent with
prior observations of incorrect generated
assertions~\citep{yuan2024manualtests}. A bracket parser
suite includes unmatched parentheses but checks only the output count,
accepting a candidate that discards valid groups. Both the poker and bracket
suites fail on the initial program and pass on the reference, yet accept an
invalid alternative. These paired outcomes show how missing input
combinations and weak assertions can survive reference evaluation while
additional patches expose them.
\textbf{Execution logic errors.} Errors in the tests' own execution can
also prevent them from evaluating candidate behavior. An EvoEval suite
accesses a function stored on its test class through \texttt{self},
unintentionally binding an extra argument and failing even on the reference
repair. Such failures explain broader reliability problems but do not
establish a single reference blind spot.

\begin{rqanswer}{3}
\textbf{Tests fail through coverage gaps, faulty checks, and execution logic
errors.} The analyzed tests omit relevant behavior, misjudge candidate
correctness, or fail during execution, helping explain limited reliability.
Paired evidence shows that missing input combinations and
weak assertions survive reference evaluation but are exposed by additional
invalid patches.
\end{rqanswer}

\subsection{RQ4: Evaluating TestHelix}
\label{sec:rq4-testhelix}

\textit{How do heterogeneous synthesis, peer cross validation, and recursive
self improvement contribute to joint test reliability?}

Table~\ref{tab:testhelix-rsi} compares five TestHelix configurations for
GPT 5.6 sol and Kimi K3, alongside native agent comparators,
Codex (max) and Kimi Code, respectively. Within each model group, the
TestHelix configurations share the same base model and tools and use the
same total inference cap per benchmark task, shared across pairs.
We compare single pair generation with three heterogeneous pairs followed
by random submission of one locally qualified suite, then enable peer cross
validation at $K=3$ and $K=5$. The final configuration applies RSI at $K=3$
using one shared strategy optimized with Opus~4.6 on 600 training tasks for
Reference Acceptance Rate and frozen for evaluation with both models. The other configurations
retain the initial strategy. Appendix~\ref{sec:testhelix-experiment-details}
provides the experimental controls, and
Appendix~\ref{app:testhelix-development} details the strategy optimization.

\begin{table}[!htbp]
\centering
\caption{\textbf{Native agent comparisons and TestHelix ablations.} HS denotes heterogeneous synthesis of $K$ test and repair pairs; PCV denotes peer cross validation. All scores are percentages.}
\label{tab:testhelix-rsi}
\begingroup
\footnotesize
\setlength{\tabcolsep}{2pt}
\renewcommand{\cmark}{\textcolor{green!50!black}{\ding{51}}}
\renewcommand{\xmark}{\textcolor{red}{\ding{55}}}
\begin{tabularx}{\linewidth}{c*{12}{C}}
\toprule[1pt]
\multicolumn{3}{c}{\textbf{Setting}}
& \multicolumn{5}{c}{\textbf{GPT 5.6 sol}}
& \multicolumn{5}{c}{\textbf{Kimi K3}} \\
\cmidrule(lr){1-3}\cmidrule(lr){4-8}\cmidrule(l){9-13}
HS ($K$) & PCV & RSI
& RAR & VAR & IRR & $H$ & \textbf{JSF}
& RAR & VAR & IRR & $H$ & \textbf{JSF} \\
\midrule
\multicolumn{3}{l}{Native agent}
& 54.00 & 61.33 & 86.00 & 71.60 & 23.33
& 56.67 & 63.67 & \textbf{83.00} & \textbf{72.06} & 24.67 \\
\midrule
1 & \xmark & \xmark
& 51.00 & 57.50 & 86.73 & 69.15 & 20.33
& 53.67 & 56.25 & 70.93 & 62.74 & 21.67 \\
3 & \xmark & \xmark
& 51.33 & 58.08 & \textbf{87.27} & 69.74 & 21.00
& 54.00 & 57.00 & 71.33 & 63.36 & 22.67 \\
3 & \cmark & \xmark
& 57.67 & 64.25 & 85.33 & 73.30 & 27.67
& 58.00 & 60.08 & 76.80 & 67.42 & 29.00 \\
5 & \cmark & \xmark
& 57.33 & 65.25 & 85.73 & 74.10 & 28.33
& 57.33 & 61.50 & 76.27 & 68.09 & 29.67 \\
\rowcolor{answerblue}
3 & \cmark & \cmark
& \textbf{60.67} & \textbf{69.17} & \textbf{87.27} & \textbf{77.17} & \textbf{32.33}
& \textbf{62.00} & \textbf{64.25} & 78.67 & 70.73 & \textbf{33.33} \\
\bottomrule[1pt]
\end{tabularx}
\endgroup
\end{table}

Across both models, larger gains come from using the generated pairs to
improve tests than from heterogeneous synthesis without peer cross validation. Generating
three heterogeneous pairs and randomly submitting one improves Joint Success Function by at
most one percentage point over single pair generation. Enabling the
complete peer cross validation process at the same $K=3$ adds more than six
points, supporting the use of alternative repair hypotheses to challenge,
refine, and select tests before submission. Expanding peer cross
validation to five pairs yields only 0.67 additional points, whereas RSI
improves the three pair process by 4.33 to 4.67 points. Under these inference
caps, improving how tests are generated and checked therefore offers a
larger gain than allocating the budget to additional pairs. The evolved
configuration achieves the highest Joint Success Function for both models and exceeds their
native coding agents, suggesting that the generation workflow remains a
useful target for improving reliability with a fixed base model.

\begin{rqanswer}{4}
\textbf{Heterogeneous synthesis helps tests generalize across implementations.}
Peer cross validation uses diverse repairs to check and correct tests,
while RSI adapts test generation strategies across tasks. This process improves
Joint Success Function over Single by 11.67 to 12.00 percentage points under matched inference
budget caps, with gains from each component in the ablations.
\end{rqanswer}

\section{Related Work}

\parabf{Coding agents and test generation benchmarks.}
SWE bench evaluates repository changes with executable
tests~\citep{jimenez2024swebench,yang2024swebenchmultimodal}; Terminal Bench
assesses command line tasks~\citep{merrill2026terminalbench}, and ContextBench
evaluates context retrieval~\citep{li2026contextbench}. SWT Bench studies tests
derived from GitHub issues~\citep{mundler2024swtbench}; RESTestBench evaluates
requirement based REST API tests using related mutants~\citep{kogler2026restestbench}.
TestPrism evaluates test generation across heterogeneous sources using fixed
valid and invalid implementation panels.
TestHelix uses AutoSaddler~\citep{park2026autosaddler} to search a large
semantic space of test generation strategies, drawing on
Meta-Harness~\citep{lee2026metaharness} and prior work on reflective feedback
and prompt optimization~\citep{shinn2023reflexion,khattab2023dspy,yuksekgonul2024textgrad,agrawal2025gepa}.

\parabf{Test generation and coverage.}
Randoop and EvoSuite use random generation and
search~\citep{pacheco2007randoop,fraser2011evosuite}. Language model approaches
evaluate execution and coverage~\citep{schafer2024llmtests,xu2025swecompass}.
CodaMOSA and CoverUp use coverage feedback to guide
generation~\citep{lemieux2023codamosa,pizzorno2025coverup}, while SymPrompt
structures prompts around execution paths~\citep{ryan2024symprompt}.
TestExplora studies repository test generation and Entry
Coverage~\citep{liu2026testexplora}. TestPrism reports task scoped coverage
alongside behavioral discrimination.

\parabf{Test adequacy and executable oracles.}
Mutation testing assesses rejection of faulty variants~\citep{jia2011mutation}.
PROBE augments repair benchmark tests with reference variants and checks
robustness under behavior preserving transformations~\citep{li2026probe}.
Joint Success Function jointly requires acceptance of valid and rejection of invalid candidates.
The oracle problem and patch plausibility limit executable
verdicts~\citep{barr2015oracle,qi2015patchplausibility}: broader observed
behavior does not establish label completeness or semantic correctness.
CodeT selects code through agreement with generated
tests~\citep{chen2022codet}; CodeRL combines test feedback with a learned
critic to refine code~\citep{le2022coderl}.

\section{Conclusion}

We introduced TestPrism to evaluate generated tests beyond a single reference.
Its Joint Success Function checks acceptance of valid alternatives and rejection
of defective implementations across diverse candidate sets. Our evaluation
reveals persistent limitations in frontier coding agents; analysis of 286
retained failed trajectories across six model families identifies coverage gaps,
faulty checks, and execution logic errors. TestHelix improves joint success
through heterogeneous synthesis, peer cross validation, and recursive self
improvement, supporting more reliable test generation across implementations.

\label{sec:main-end}
\clearpage
\endgroup
\bibliography{references}
\bibliographystyle{iclr2027_conference}

\clearpage
\appendix
\renewcommand{\parabf}[1]{\noindent\textbf{#1}\enspace}

\subsection*{Appendix Guide}

The appendices follow the order of the main text. Each entry below links
to the corresponding supporting material.

\begin{tcolorbox}[
  colback=promptback,
  colframe=promptborder,
  boxrule=0.45pt,
  arc=2.5pt,
  left=8pt,right=8pt,top=6pt,bottom=6pt]
\small
\begin{tabularx}{\linewidth}{@{}l X r@{}}
\textbf{Main text} & \textbf{Appendix} & \textbf{Page} \\
\midrule
Introduction & \hyperref[sec:benchmark-comparison-details]{\ref*{sec:benchmark-comparison-details}. Benchmark comparison} & \pageref{sec:benchmark-comparison-details} \\
TestPrism & \hyperref[app:testprism-details]{\ref*{app:testprism-details}. Construction and evaluation} & \pageref{app:testprism-details} \\
TestHelix & \hyperref[app:testhelix-details]{\ref*{app:testhelix-details}. Method and strategy details} & \pageref{app:testhelix-details} \\
Experiments & \hyperref[sec:experimental-details]{\ref*{sec:experimental-details}. Results and analysis} & \pageref{sec:experimental-details} \\
Scope & \hyperref[sec:limitations]{\ref*{sec:limitations}. Limitations} & \pageref{sec:limitations} \\
\end{tabularx}

\noindent\textbf{Research questions:}
\hyperref[app:rq1-details]{RQ1~(\ref*{app:rq1-details})}\quad
\hyperref[sec:patch-count-diagnostic]{RQ2~(\ref*{sec:patch-count-diagnostic})}\quad
\hyperref[sec:error-taxonomy-details]{RQ3~(\ref*{sec:error-taxonomy-details})}\quad
\hyperref[sec:testhelix-experiment-details]{RQ4~(\ref*{sec:testhelix-experiment-details})}.
\end{tcolorbox}

\section{Benchmark Comparison Conventions}
\label{sec:benchmark-comparison-details}

Table~\ref{tab:compare_benches} compares behavioral and coverage targets:
a check marks the corresponding target, a dagger marks a related check with
different granularity or definitions, and a cross marks its absence from the
main protocol. Reference Acceptance Rate measures the fail to pass transition
from the initial program to the reference implementation.

\paragraph{Behavioral and coverage targets.}
TestGenEval uses mutation testing; CLOVER and MultiFileTest evaluate tests
through execution and code coverage
\citep{jain2025testgeneval,xu2025clover,wang2026multifiletest}.
Their file, block, or project coverage is distinct from reference change and
entry scope coverage. SWT Bench and TDD Bench Verified measure changed code
coverage \citep{mundler2024swtbench,ahmed2024tddbench}.
OmniCode jointly requires reference acceptance and rejection of all bad
patches, while SWE Mutation reports mutant detection
\citep{omnicode2026,swemutation2026}. These checks provide evidence beyond one
reference; acceptance of valid alternatives under Joint Success Function is
outside their scope.

\paragraph{Counting conventions.}
Variants counts implementations other than the reference, including mutants;
a check indicates that variants are used but no total is established. SWT
Bench's separate patch filtering experiment is outside that column's main
protocol scope. TestPrism has 2,700 such variants and 300 reference solutions.
Languages counts individual task implementation languages. The 300 TestPrism
tasks use Python (205), TypeScript (44), Java (19), JavaScript (16), C++ (8),
and Go (8). Scope identifies test granularity; TestPrism spans local function
and file tests, multifile projects, and repository issues.

\clearpage
\section{TestPrism: Construction and Evaluation}
\label{app:testprism-details}

This appendix follows the construction steps, benchmark statistics, and
evaluation protocol in Section~\ref{sec:testprism}.

\label{app:construction-details}

\subsection{Task Screening and Adaptation}

\parabf{Source selection and prior evidence.}
We select public benchmarks with traceable task origins and documented test
validation. The \href{https://www.swebench.com/}{SWE bench} and
\href{https://www.tbench.ai/leaderboard/terminal-bench/2.0}{Terminal Bench}
families maintain public leaderboards with results from multiple agent developers.
Other sources draw on community maintained Exercism exercises, official
USACO competitions, and published benchmarks for specialized programming
settings.

\parabf{Task adaptation.}
We retain tasks centered on code and suitable for executable test authoring,
excluding tasks with unusable Docker images or environments that cannot be
reproduced. Adapters map tasks from all \benchSourceCount{} source benchmarks
to Harbor's instruction, environment, tests, and solution components.
Dependencies, services, and execution limits follow each source so that its
original functional and performance requirements can be checked.

\subsection{Task Selection}
\label{app:task-selection}

\paragraph{Test set selection and eligibility.}
Initial screening excludes tasks outside the code testing scope and tasks
whose environments cannot be reproduced.
The selection signals in Section~\ref{sec:selection} use effective lines in
\texttt{solution.sh}, excluding blank and comment lines
(\texttt{solution\_lines}); test functions enumerated from source files or
benchmark test metadata (\texttt{test\_func\_count}); and this count divided
by the number of detected modified files, with a minimum denominator of one
(\texttt{test\_coverage\_ratio}). The ratio describes test density.
We apply source-specific lower bounds calibrated to each source's
distribution. Sources with smaller typical solutions or lower enumerated
test counts use correspondingly lower bounds, preserving their representation
alongside sources with larger values. This accounts for differences in task
granularity and verifier structure; source quotas additionally balance the
benchmark composition.
After enrichment, final admission requires applicability annotations for both
$\Delta\mathcal{C}$ and Entry Coverage, ten physical candidate artifacts meeting
the diversity criteria in Appendix~\ref{app:diversity-conversion}, and
agreement between each artifact digest and its manifest.

Table~\ref{tab:source-selection-bounds} lists the source-specific thresholds
applied during difficulty screening. Source test counts follow the complete and partial
enumeration convention in Appendix~\ref{app:sources}. Density uses original
patch paths for 204 tasks and reference file changes reconstructed from
initial task files and build specifications for the remaining 96. It counts
distinct changed, added, or removed files, including code, configuration,
tests, and documentation; unchanged rewrites, temporary helpers, and runtime
or build outputs are excluded.

\begin{table}[htbp]
\centering
\caption{\textbf{Source-specific filtering thresholds.} Task counts describe the final test set.}
\label{tab:source-selection-bounds}
\begingroup
\footnotesize
\setlength{\tabcolsep}{4pt}
\begin{tabularx}{\linewidth}{@{}Xrrrr@{}}
\toprule
Source & Tasks & Lines & Tests & Density \\
\midrule
Aider Polyglot & 8 & 32 & 20 & 20 \\
AlgoTune & 2 & 73 & 2 & 0.5 \\
DevEval & 12 & 7 & 2 & 0.2 \\
EvoEval & 20 & 29 & 2 & 2 \\
Long CLI & 1 & 411 & 28 & $28/3$ \\
QuixBugs & 16 & 36 & 2 & 2 \\
SWE bench & 33 & 11 & 9 & $13/3$ \\
SWE bench Live & 28 & 14 & 7,424 & $3871/3$ \\
SWE bench Multimodal & 11 & 21 & 30 & 2 \\
SWE Perf & 21 & 32 & 3 & 0.6 \\
SWE rebench & 46 & 17 & 16 & $4/15$ \\
SWE bench Pro & 17 & 669 & 1 & $1/15$ \\
SWE PolyBench & 48 & 18 & 11 & 1.375 \\
Terminal Bench & 8 & 83 & 4 & 4 \\
Terminal Bench 2 & 4 & 15 & 6 & 6 \\
Terminal Bench Pro & 5 & 66 & 12 & 6 \\
USACO & 20 & 9 & 2 & 2 \\
\bottomrule
\end{tabularx}
\endgroup
\end{table}

\paragraph{Training and test split.}
Tasks are normalized across sources and deduplicated before partitioning.
Difficulty screening operates within source benchmarks. Source quotas form
a training set of 1,700 tasks and a disjoint candidate test pool. The training set is
fixed at this stage; candidate test tasks continue through verifier validation,
patch generation, and diversity filtering. We select the final
\benchTaskCount{} eligible test tasks with proportional source quotas,
retaining every source, and seed 42 for sampling within each source.
Tasks not selected for the final test set are excluded from its statistics.
Candidate test tasks undergo the verifier checks and independent human
review described in Appendix~\ref{app:source-test-validation} before final
admission.

Selection uses task properties and artifact availability, independently of
Reference Acceptance Rate, Joint Success Function, or coverage collection success. Test set
membership remains fixed when runtime observations are missing.

\parabf{Filtering funnel.}
Table~\ref{tab:construction-funnel} summarizes construction from an
initial pool of 15,148 source tasks. The final dataset contains 1,700 training tasks and
\benchTaskCount{} test tasks. All entries count tasks, including stages
that assess their patch sets. Source quota exclusions denote tasks left
unselected, rather than failures of the quality checks. After the split,
the training set remains fixed and only the candidate test pool continues through
verifier validation, initial panel assembly, and diversity refinement.
Initial assembly requires one reference, four valid alternatives, and five
invalid implementations. Tasks that exhaust the shared generation budget
before filling these quotas fail initial assembly. Tasks that fill the
quotas but still fail either group criterion when the same budget is
exhausted are excluded during diversity refinement
(Appendix~\ref{app:diversity-conversion}).

\begin{table}[htbp]
\centering
\caption{\textbf{Task filtering stages in TestPrism construction.}}
\label{tab:construction-funnel}
\begingroup
\small
\setlength{\tabcolsep}{4pt}
\begin{tabularx}{\linewidth}{@{}Xrrr@{}}
\toprule[0.7pt]
\textbf{Stage} & \textbf{Retained} & \textbf{Excluded} & \textbf{Rate} \\
\midrule
\multicolumn{4}{@{}l}{\textit{Shared screening}} \\
Code focus and test authoring suitability & 12,937 & 2,211 & 14.6\% \\
Docker and environment reproducibility & 10,984 & 1,953 & 15.1\% \\
Solution size, test count, and test density & 6,731 & 4,253 & 38.7\% \\
Source quota selection & 2,473 & 4,258 & 63.3\% \\
\midrule
\multicolumn{4}{@{}l}{\textit{Split: 1,700 training tasks and 773 candidate test tasks}} \\
\midrule
\multicolumn{4}{@{}l}{\textit{Candidate test pool only}} \\
Verifier checks and human review & 608 & 165 & 21.3\% \\
Initial candidate panel assembly & 394 & 214 & 35.2\% \\
Within-group diversity refinement & 311 & 83 & 21.1\% \\
Final source quota selection & 300 & 11 & 3.5\% \\
\bottomrule[0.7pt]
\end{tabularx}
\endgroup
\end{table}

\FloatBarrier

\subsection{Patch Generation and Validation}

\subsubsection{Source Test Validation}
\label{app:source-test-validation}

Table~\ref{tab:source-test-validation} links all
\benchSourceCount{} sources to their test provenance and original quality
controls, including execution checks, expert review, and validation against
reference or alternative solutions.

\parabf{Additional validation after adaptation.}
The source benchmarks provide the original validation; our additional
validation round checks the adapted verifiers before candidate generation.
We preserve the source verifiers during adaptation and require them to
reject the initial state and accept the reference solution before using
them to label candidates. For performance tasks, failure means the initial,
unaccelerated program misses the task's performance requirement; passing
requires an accelerated solution that meets the original task-defined
thresholds while preserving functional correctness.
Candidate labels are established independently
of the generated tests under evaluation and remain fixed across agents.

\parabf{Independent verifier review.}
Before candidate generation, two experts independently review the task
requirements and source verifier, judging whether the available information
supports test authoring. Drawing on development experience, they assess
potential overfitting to the reference and gaps in requirement coverage,
including checks that could reject valid alternatives or miss incorrect
behavior.
Disagreements are discussed to clarify the concerns; any task with a
remaining concern about verifier usability is excluded. Agreement on the
independent usability judgments, before adjudication, is Cohen's
$\kappa=0.68$. This review complements the source validation procedures and
automatic execution checks; retained candidate labels follow the verifier
outcomes defined in Section~\ref{sec:panels}.

\begin{table}[!htbp]
\centering
\footnotesize
\setlength{\tabcolsep}{4pt}
\caption{\textbf{Source test provenance and prior validation evidence.}}
\label{tab:source-test-validation}
\begin{tabularx}{\linewidth}{@{}>{\raggedright\arraybackslash}p{0.19\linewidth}>{\raggedright\arraybackslash}p{0.25\linewidth}>{\raggedright\arraybackslash}X@{}}
\toprule
Source & Test provenance & Original quality controls \\
\midrule
\href{https://github.com/Aider-AI/polyglot-benchmark}{Aider Polyglot} & Community maintained Exercism suites & Expected outputs, example implementations, and reviewed \href{https://github.com/exercism/problem-specifications}{canonical specifications} support specification checks and traceability. \\
AlgoTune & Expert solvers, input generators, and verifiers & Two additional contributors review each task; outputs from different solver seeds check acceptance of alternative solutions~\citep{press2025algotune}. \\
DevEval (DevBench) & Project unit and acceptance tests & Annotators execute reference code and supplement existing tests to improve coverage~\citep{li2024devbench}. \\
EvoEval & Generated tasks, solutions, and augmented tests & Authors inspect tasks and reference solutions; iterative solution agreement resolves ambiguities~\citep{xia2024evoeval}. \\
Long CLI & Tests written from requirements alone & Tests are authored without the solution repository; initial and solution states undergo failure and regression checks followed by expert review~\citep{feng2026longcli}. \\
QuixBugs & Quixey challenge programs with passing and failing cases & Released drivers compare buggy and corrected programs on supplied cases~\citep{lin2017quixbugs}. \\
SWE bench & Developer tests from merged pull requests & Execution requires a fail to pass transition; installation and runtime failures are filtered~\citep{jimenez2024swebench}. \\
SWE bench Live & Tests from issue resolving pull requests & Validation requires a fail to pass transition and consistent results across repeated runs~\citep{zhang2025swebenchlive}. \\
SWE bench Multimodal & Repository tests for visual software issues & Ten validation runs remove inconsistent tests; authors inspect every task for feasibility~\citep{yang2024swebenchmultimodal}. \\
SWE Perf & Repository tests for performance changes & Selected tests pass before and after optimization; timing validation uses 20 repetitions and statistical filtering~\citep{he2025sweperf}. \\
SWE rebench & Test patches from resolving pull requests & Reference patches must fix initial test failures while preserving initially passing tests~\citep{badertdinov2025swerebench}. \\
SWE bench Pro & Repository tests with human review & Repeated runs check stability; human reviewers remove irrelevant or overly broad tests~\citep{deng2025swebenchpro}. \\
SWE PolyBench & Repository tests and pull request updates & Manually configured containers check failure transitions and regression preservation, excluding untestable changes~\citep{rashid2025swepolybench}. \\
Terminal Bench & Contributor tests and reference solutions & \href{https://github.com/harbor-framework/terminal-bench-1/tree/d28711d0da2675d0bb1d56de45ae5df6082438a3/.github}{Contribution checks} cover specification alignment, reference success, and failure without agent actions. \\
Terminal Bench 2 & Contributor tests and manual reference solutions & Three reviewers check specification, execution, and exploits, averaging three review hours per task~\citep{merrill2026terminalbench}. \\
Terminal Bench Pro & Programmer authored instructions and unit tests & Independent experts review ambiguity, false positives, reference solutions, and coverage over multiple rounds~\citep{wang2025rome}. \\
USACO & Official hidden tests and editorial based reference code & Reference solutions pass hidden tests; manual resource checks compare correct and incorrect asymptotic complexity~\citep{shi2024olympiad}. \\
\bottomrule
\end{tabularx}
\end{table}

\subsubsection{Candidate Generation and Labeling}

Only the candidate test pool undergoes patch generation, using different
coding agents and models of varying capabilities to produce alternative
implementations. These runs capture multiple ways of satisfying the task
requirements and errors arising during task solving. Each candidate is
stored as a patch artifact. The admission gate removes empty submissions and
exact duplicates, then checks applicability in the adapted task environment.
The source verifier executes its checks in a clean environment, producing a
matrix of candidate outcomes. A candidate is valid when every check passes
and invalid when any check fails, regardless of its intended correctness.
We retain its provenance, identifier, label, and digest for selection and
evaluation. Each task has a single budget of 30 generated patch outputs,
shared by initial panel assembly and diversity refinement. The existing
reference does not consume this budget. Every produced patch counts once,
including patches later rejected or not selected. No per-label generation
cap or separate refinement-round cap applies.

\subsubsection{Candidate Patch Set}
\label{app:candidate-patches}

Table~\ref{tab:benchmark300-corpus} summarizes all \benchPatchCount{}
candidate artifacts belonging to the fixed task manifest.
Each task contributes one reference implementation, four other valid
implementations, and five invalid implementations. Their executable labels
and digests are retained together. Every referenced file is present and
matches its manifest digest. Candidate labels follow the validated source
verifiers described in Appendix~\ref{app:source-test-validation}.
Panel diversity is assessed using the criteria for valid and invalid groups
in Appendix~\ref{app:diversity-conversion}.

\begin{table}[htbp]
\centering
\small
\setlength{\tabcolsep}{4pt}
\caption{\textbf{Candidate patch inventory for the fixed TestPrism test set.}}
\label{tab:benchmark300-corpus}
\begin{tabular}{rrrrrr}
\toprule
Tasks & Reference & Other valid & Invalid & Verified files & Missing \\
\midrule
300 & 300 & 1200 & 1500 & 3000 & 0 \\
\bottomrule
\end{tabular}
\end{table}

\FloatBarrier

\subsection{Diversity Filtering and Task Conversion}
\label{app:diversity-conversion}

\parabf{AST distance.}
We apply each patch to the same initial program and parse the modified
production code before and after the change with fixed versions of
Tree-sitter and the corresponding language grammars. From AST nodes in the
modified regions on each side, we extract ordered local subtrees of depth
three, counting the root as depth zero. Features preserve node types, child
order, operators, and literal types and values.
Identifiers are normalized to a common token; comments, formatting, and
source positions do not contribute features.

Let $B_p$ and $C_p$ be the resulting feature sets in the initial and candidate
states. We represent the patch by its added and removed structural features:
\[
\mathcal{A}(p)=
\bigl(\{+\}\times(C_p\setminus B_p)\bigr)
\cup
\bigl(\{-\}\times(B_p\setminus C_p)\bigr).
\]
Each set depends only on the patch and the common initial program, so its
definition remains fixed across comparisons. Features are deduplicated;
repeated occurrences do not increase their weight. We use Jaccard distance:
\[
d_{\mathrm{AST}}(p,q)
=1-\frac{|\mathcal{A}(p)\cap\mathcal{A}(q)|}
{|\mathcal{A}(p)\cup\mathcal{A}(q)|}.
\]
When both sets are empty, the distance is zero; when only one is empty, it
is one. Unresolved parse errors in the changed regions leave the distance
unavailable. Such a generated candidate cannot enter the selected panel and
still counts toward the shared 30-patch generation budget. This measure captures local
structural differences; verifier outcomes determine behavioral labels.

\parabf{Group admission criteria.}
We first assemble a valid group $V=\{p_{\mathrm{ref}},v_1,v_2,v_3,v_4\}$
and an invalid group $I=\{i_1,i_2,i_3,i_4,i_5\}$.
Each group contains ten distinct unordered pairs. The valid group must satisfy
\[
D(V)=\frac{1}{10}\sum_{\{p,q\}\subset V}d_{\mathrm{AST}}(p,q)>0.5.
\]
This mean includes comparisons with the reference and between alternatives;
it is not a lower bound on every pair.

For invalid implementations, we first compare the sets of verifier checks
they fail. Different failed-check sets satisfy the diversity check directly.
Candidates with the same failed-check set must additionally pass the AST
Jaccard check. Thus, AST filtering within the invalid group is conditional
on matching failed-check sets.
We impose no cross-group distance constraint: verifier outcomes
already distinguish implementations that pass all checks from those that
fail at least one.

\parabf{Generation and replacement.}
If either group fails its criterion, we continue generating candidates.
Each new candidate undergoes the same applicability, deduplication, and
verifier checks. Its observed label determines the group it can enter.
When the valid group has not met its criterion, we evaluate each replacement
of a nonreference member and choose the one that maximizes $D(V)$, accepting
it only if the mean strictly increases. For the invalid group, replacements
are assessed by their failed-check sets, with the AST Jaccard check applied
when those sets match.
The reference remains fixed, and the panel always contains one reference,
four valid alternatives, and five invalid implementations. We stop as soon
as both criteria hold or after evaluating the thirtieth generated patch.
At that limit, tasks without a complete panel satisfying both criteria are
excluded. Final source quotas select from the eligible tasks; tasks left
unselected at this stage have passed the quality filters. Every retained
artifact must match its recorded digest.

\parabf{Test generation adapter.}
Conversion primarily updates \texttt{task.yaml} to specify submission
requirements and \texttt{Dockerfile} to prepare the execution environment.
The original requirement remains the behavioral contract, and candidate
patches are reserved for evaluation. Submitted tests, fixtures, and the
executable entry script reside in the designated artifact directory.

\parabf{Interpretation boundaries.}
The source and difficulty composition reflects the development collection.
Requiring meaningful local production targets for $\Delta\mathcal{C}$ and
Entry Coverage excludes tasks whose essential outcomes lack such targets.
Candidate labels follow the validated source task contracts, while
panel diversity follows the criteria for valid and invalid groups above.
Runtime and coverage observation counts are reported separately in
Appendix~\ref{sec:evaluation-details}.

\FloatBarrier

\FloatBarrier
\subsection{Benchmark Statistics}
\label{app:sources}

\begin{figure}[htbp]
\centering
\includegraphics[width=0.48\linewidth]{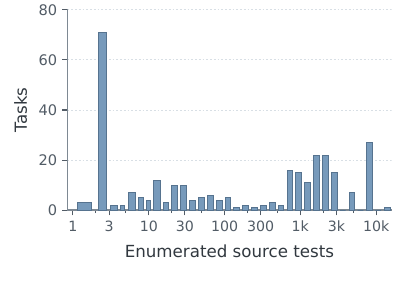}
\caption{\textbf{Distribution of source test counts.}}
\label{fig:test-count-distribution}
\end{figure}

\begin{figure}[htbp]
\centering
\includegraphics[width=\linewidth]{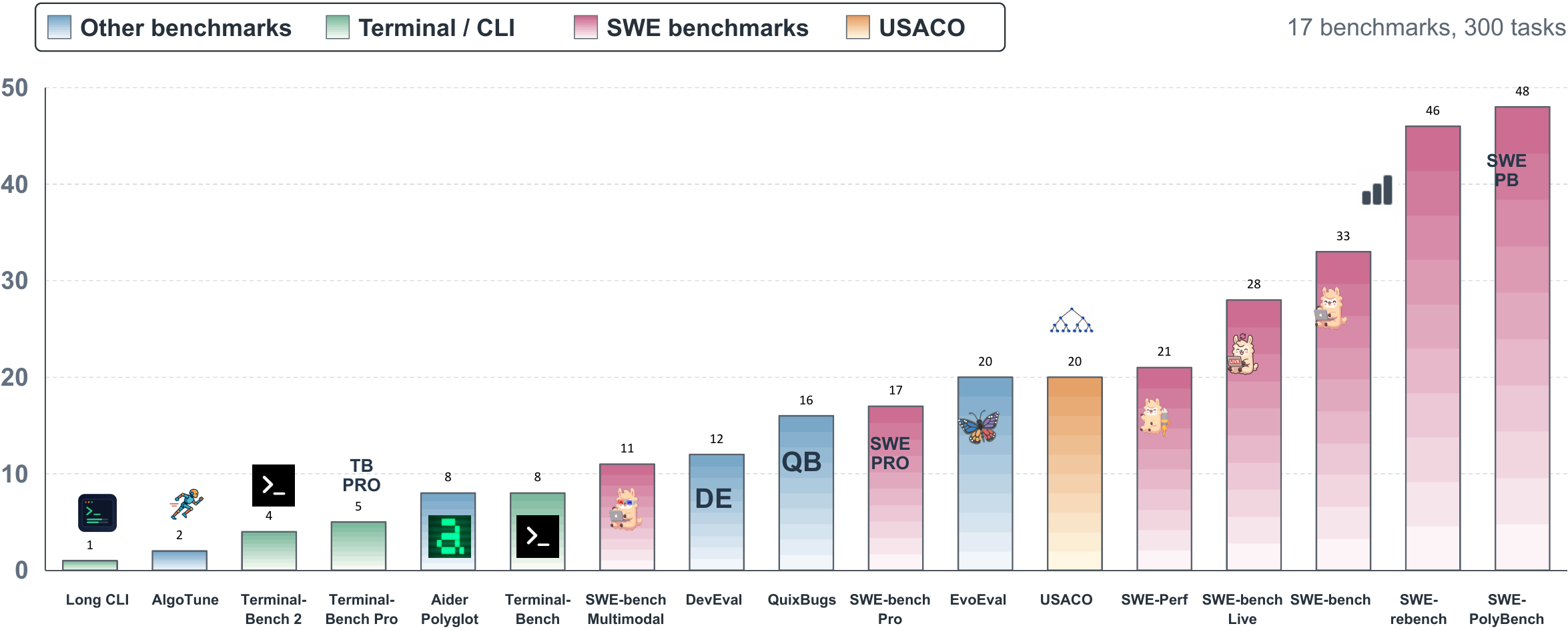}
\caption{\textbf{Test tasks by source benchmark.}}
\label{fig:benchmark-statistics}
\end{figure}

\begin{figure}[htbp]
\centering
\includegraphics[width=\linewidth]{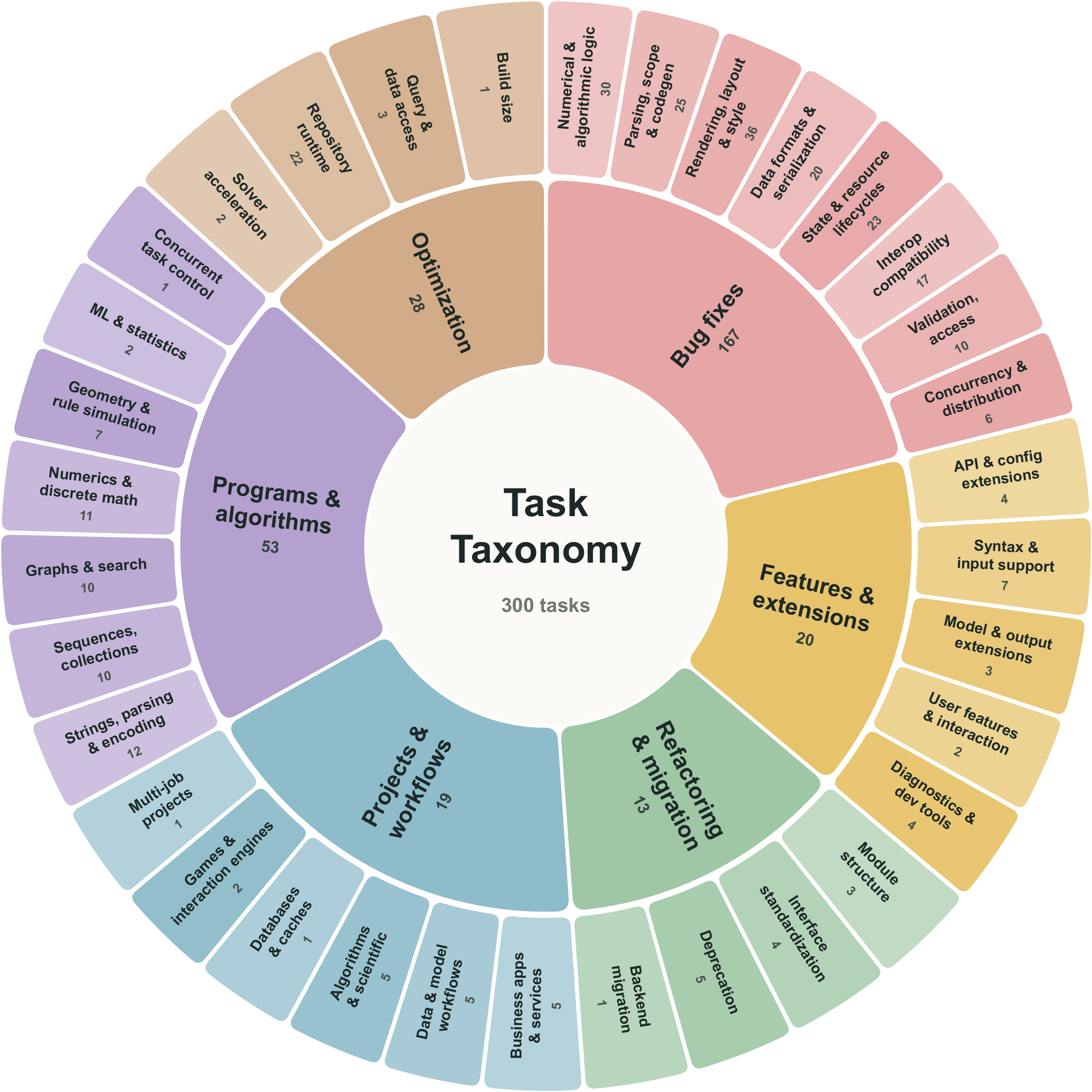}
\caption{\textbf{Task taxonomy of TestPrism.} Numbers indicate task counts;
sector angles and areas are not proportional to these counts.}
\label{fig:task-taxonomy}
\end{figure}

\parabf{Reading the statistics.}
Figure~\ref{fig:benchmark-statistics} reports task counts by source benchmark;
Figure~\ref{fig:task-taxonomy} reports category and subcategory counts.
Figure~\ref{fig:test-count-distribution} shows the distribution of enumerated
source test counts.
Table~\ref{tab:language-statistics} gives test set task, repository, source
test, and patch counts by language, with counting conventions below.
Unselected tasks in the candidate test pool do not enter these statistics.

Table~\ref{tab:benchmark300-sources} and Figure~\ref{fig:benchmark-statistics}
report the task sources in the test set.
The counts cover all \benchTaskCount{} selected tasks, with all
\benchSourceCount{} sources retained.
Source names identify provenance; each task's requirement and coverage
applicability annotations determine its measurement scope.
Figure~\ref{fig:task-taxonomy} shows the six task categories, 34 subcategories,
and their test task counts.

\begin{table}[htbp]
\centering
\footnotesize
\setlength{\tabcolsep}{3pt}
\caption{\textbf{Test task counts by source benchmark.}}
\label{tab:benchmark300-sources}
\begin{tabularx}{\linewidth}{l*{6}{>{\centering\arraybackslash}X}}
\toprule
Source & Aider Polyglot & AlgoTune & DevEval & EvoEval & Long CLI & QuixBugs \\
Tasks & 8 & 2 & 12 & 20 & 1 & 16 \\
\midrule
Source & SWE bench & SWE bench Live & SWE bench Multimodal & SWE Perf & SWE rebench & SWE bench Pro \\
Tasks & 33 & 28 & 11 & 21 & 46 & 17 \\
\midrule
Source & SWE PolyBench & Terminal Bench & Terminal Bench 2 & Terminal Bench Pro & USACO &  \\
Tasks & 48 & 8 & 4 & 5 & 20 &  \\
\bottomrule
\end{tabularx}
\end{table}

\parabf{Language and source test counts.}
Table~\ref{tab:language-statistics} reports test set task, repository, source
test, and patch counts by language.
Repository counts cover distinct source repository and project
identities recorded in task files, with totals deduplicated across languages.
Source test counts enumerate test identifiers and verifier checks from those
files, summed across tasks. When inner cases are enumerated, their aggregate
wrapper is not counted again. Test patch counts come from the fixed
candidate inventory and include reference solutions, for a total of 3,000.

\begin{table}[htbp]
\centering
\caption{\textbf{TestPrism test set statistics by language.} \#Test counts enumerated source tests and verifier checks, including partial counts.}
\label{tab:language-statistics}
\begingroup
\small
\setlength{\tabcolsep}{3pt}
\begin{tabularx}{\linewidth}{@{}l*{4}{>{\raggedleft\arraybackslash}X}@{}}
\toprule[0.7pt]
\textbf{Language} & \textbf{\#Repo} & \textbf{\#Task}
 & \textbf{\#Test} & \textbf{\#Patch} \\
\midrule
Python & 46 & 205 & 327,627 & 2,050 \\
JS/TS & 9 & 60 & 110,954 & 600 \\
Java & 4 & 19 & 38 & 190 \\
C/C++ & 5 & 8 & 50 & 80 \\
Go & 3 & 8 & 144 & 80 \\
\rowcolor{answerblue}
\textbf{Total} & 67 & 300 & 438,813 & 3,000 \\
\bottomrule[0.7pt]
\end{tabularx}
\endgroup

\end{table}

Figure~\ref{fig:test-count-distribution} uses the same task input inventory:
184 tasks have complete test lists and 116 have partial counts from
enumerable verifier functions or performance checks. All 300 tasks have
positive enumerable counts; unenumerated internal tests are excluded.

Language labels follow test entry annotations. JS/TS combines JavaScript
and TypeScript, and C/C++ combines C and C++.

\FloatBarrier

\FloatBarrier
\subsection{Evaluation Protocol}
\label{sec:evaluation-details}

This section expands the auxiliary definitions and collection conventions
for the evaluation protocol in Section~\ref{sec:evaluation-protocol}.

\subsubsection{Behavioral Metric Definitions}
\label{sec:behavioral-details}

For task $i$, let $o(T_i,p)$ denote a recorded outcome in
$\{\mathrm{pass},\mathrm{fail},\mathrm{error}\}$ when the submitted entry
script $T_i$ runs on implementation $p$. The acceptance and rejection indicators are
\[
a_i(p)=\mathbf{1}\bigl[o(T_i,p)=\mathrm{pass}\bigr],
\qquad
r_i(p)=\mathbf{1}\bigl[o(T_i,p)=\mathrm{fail}\bigr].
\]
Both indicators are zero when the outcome is missing or a recognized execution
error. The reference transition and full panel success indicators are
\[
F_i=r_i(b_i)\,a_i(g_i),
\qquad
S_i=F_i\,\prod_{p\in\mathcal{V}_i\setminus\{g_i\}} a_i(p)\,
          \prod_{p\in\mathcal{I}_i} r_i(p).
\]
Since $g_i\in\mathcal{V}_i$, substituting $F_i$ into $S_i$ gives the joint
criterion in the main text. The benchmark rates average these indicators over
the fixed set of $N=\benchTaskCount{}$ tasks:
\[
\text{Reference Acceptance Rate}=\frac{1}{N}\sum_{i=1}^{N}F_i,
\qquad
\text{Joint Success Function}=\frac{1}{N}\sum_{i=1}^{N}S_i.
\]

Let $\mathcal{V}^{\mathrm{obs}}_i$ contain the other valid candidates with
recorded outcomes, excluding $g_i$, and $\mathcal{I}^{\mathrm{obs}}_i$ the
invalid candidates with recorded outcomes. The corresponding task rates are
\[
A_i=\frac{\sum_{p\in\mathcal{V}^{\mathrm{obs}}_i}a_i(p)}
          {\lvert\mathcal{V}^{\mathrm{obs}}_i\rvert},
\qquad
R_i=\frac{\sum_{p\in\mathcal{I}^{\mathrm{obs}}_i}r_i(p)}
          {\lvert\mathcal{I}^{\mathrm{obs}}_i\rvert}.
\]
Complete observations give denominators of four for $A_i$ and five for $R_i$;
an empty observed denominator leaves the corresponding task rate undefined.
Let $\mathcal{T}_V=\{i:\lvert\mathcal{V}^{\mathrm{obs}}_i\rvert>0\}$
and $\mathcal{T}_I=\{i:\lvert\mathcal{I}^{\mathrm{obs}}_i\rvert>0\}$.
The reported component rates average the defined task rates with equal weight:
\[
\text{Valid Acceptance Rate}=\frac{1}{\lvert\mathcal{T}_V\rvert}
  \sum_{i\in\mathcal{T}_V}A_i,
\qquad
\text{Invalid Rejection Rate}=\frac{1}{\lvert\mathcal{T}_I\rvert}
  \sum_{i\in\mathcal{T}_I}R_i.
\]
Neither rate requires a successful initial-to-reference transition.
Candidates without an execution record are
excluded from the corresponding task denominator. Recorded errors remain
in it even if execution did not complete, but contribute to neither numerator.
A task with an undefined rate is excluded from that metric's mean; an empty
task set leaves the aggregate undefined. With complete observations, both
task sets contain all $N$ tasks. When every included task has the same
candidate count, the task mean equals the ratio of total matching outcomes
to total candidate observations. Different task denominators can make
these two averages differ.

Every reported baseline and TestHelix configuration has complete observations
on all 300 tasks: 1,200 valid-alternative and 1,500 invalid-candidate records.

The harmonic mean $H$ combines the aggregate acceptance and rejection rates:
\[
H=\frac{2\,\text{Valid Acceptance Rate}\,\text{Invalid Rejection Rate}}{\text{Valid Acceptance Rate}+\text{Invalid Rejection Rate}}.
\]
It is zero when both rates are zero and undefined when either rate is
undefined. We first round Valid Acceptance Rate and Invalid Rejection Rate
to the two decimal places shown as
percentages in the tables, compute their harmonic mean, and then round
$H$ to two decimal places.

The harness recognizes errors such as a missing entry script or a timeout and
otherwise uses the test runner's exit status. A nonzero exit may reflect an
assertion failure or another problem inside the script, so interpreting a
candidate disagreement requires inspecting the test and source. This limitation
also applies to a recorded failure on the initial state.

\subsubsection{Coverage Collection and Aggregation}
\label{sec:coverage-details}

We use Change Coverage from SWT Bench and Entry Coverage from
TestExplora~\citep{mundler2024swtbench,liu2026testexplora}, with measurement
defined by the execution sets below.

For $\Delta\mathcal{C}$, $C_i$ contains source coordinates changed by the
reference solution and executed by trusted tests in the corresponding initial
or reference state.
Coordinates include the program state: a replaced line can contribute one
target in each state. Generated test execution in the corresponding states
forms $H_i$. Trusted test execution fixes the target denominator $\lvert C_i\rvert$.
Changed lines without an observed executable event do not contribute to it.
The metric does not separately remove formatting edits on executed code lines.

For Entry Coverage, trusted tests identify production entry points with a unique,
auditable mapping to local source spans in the reference state. The collector
forms $E_i$ from executable coordinates in those spans, using the union of
executed and unexecuted coordinates in the generated tests' reference state
coverage report. Generated test execution on the reference state forms $H_i^g$.
Trusted tests fix the entry scope; its executable coordinates need not have
been executed by trusted tests. Overlapping entry spans count each coordinate
once, and the scope can extend beyond changed lines.

The task coverage ratios are
\[
\Delta\mathcal{C}_i = \frac{\lvert H_i\cap C_i\rvert}{\lvert C_i\rvert},
\qquad
\text{Entry Coverage}_i = \frac{\lvert H_i^{g}\cap E_i\rvert}{\lvert E_i\rvert}.
\]

Every benchmark task meets both applicability rules. An empty measured
denominator or failed collector receives an explicit status and no numeric
coverage value. With valid observations for all $N$ tasks, the benchmark
coverages are task means:
\[
\Delta\mathcal{C}=\frac{1}{N}\sum_{i=1}^{N}\Delta\mathcal{C}_i,
\qquad
\text{Entry Coverage}=\frac{1}{N}\sum_{i=1}^{N}\text{Entry Coverage}_i.
\]
All fourteen baseline configurations in Tables~\ref{tab:benchmark300-main}
and~\ref{tab:additional-baseline} have 300 valid observations for each
coverage metric, with no failed collections or empty measured denominators.

\subsubsection{Mapping Stored Metrics to Reported Quantities}

The stored \texttt{f2p\_rate} field records the single reference success
indicator $F_i$: initial failure and reference acceptance by the submitted entry
script. Reference Acceptance Rate averages this indicator over the fixed
benchmark, with missing or erroneous required outcomes receiving no success
credit.
A pass on the reference alone remains an execution record and is not a separate
reported metric.

The stored valid acceptance rate includes all five valid candidates, including
the reference. The task rate $A_i$ above covers the other four candidates for
Valid Acceptance Rate. Invalid Rejection Rate measures rejection of
the invalid candidates. Both metrics first compute each task's rate from
recorded outcomes, including recorded errors as unmatched, and then average
the defined task rates with equal weight. This follows the task averaging
convention of the stored run summaries. The stored valid acceptance rate
retains the reference, so its task rate must be recomputed over the other
four valid candidates before averaging to report Valid Acceptance Rate.
The stored balanced patch accuracy is the arithmetic mean of the stored valid
acceptance rate and invalid rejection rate. The mutation score aliases the
stored invalid rejection rate, so it provides no additional independent observation.

Joint Success Function averages the stored task resolution indicator over
the fixed benchmark.
A positive indicator requires initial failure, acceptance of every valid
candidate, rejection of every invalid candidate, and no recorded execution
error. Missing required outcomes receive no joint success credit.

Change Coverage ($\Delta\mathcal{C}$) is stored in
\texttt{gold\_patch\_coverage}; the general coverage field aliases its ratio.
Entry Coverage is stored in \texttt{entry\_coverage}.
Missing or undefined coverage values remain observation gaps rather
than numeric zeros. These correspondences are checked alongside task membership
and panel identity in the analysis artifact.

\clearpage
\section{TestHelix: Method and Strategy Details}
\label{app:testhelix-details}

This appendix follows heterogeneous synthesis, peer cross validation,
and recursive self improvement in Section~\ref{sec:testhelix}.
The component comparisons appear in
Appendix~\ref{sec:testhelix-experiment-details}.

\label{app:testhelix-policy}

This appendix documents the TestHelix strategy schema and
fixed prompt templates, followed by the RSI search configuration and
independent validation protocol.

\subsection{Heterogeneous Synthesis}
\label{app:testhelix-hs}
\label{app:testhelix-source-prompts}

\parabf{Prompt Templates.}
The following passages preserve the fixed prompt wording, including
executable names and punctuation; separate quote blocks are excerpts rather than one
continuous prompt. Task text, branch strategy, guidance, retrieved lessons,
execution observations, and remaining budget are inserted at runtime.
In these prompts, \texttt{policy} refers to the configurable test
generation strategy.

\parabf{Candidate generation.}
The candidate-generation template contains the following instructions.

\begin{promptcard}{01}{Candidate generation}
\begin{quote}\small\ttfamily\raggedright
The original public task is the acceptance contract.
\end{quote}
\promptdivider

\begin{quote}\small\ttfamily\raggedright
The unchanged baseline must fail for a task-related reason. Derive
expectations independently from public evidence, not from the repair.
Neither a command exit code nor a host checkpoint proves correctness.
\end{quote}
\promptdivider

\begin{quote}\small\ttfamily\raggedright
Only agent-tests/ at the original working directory is exported; keep
that location unchanged even if public source lives elsewhere. Keep
test scripts, fixtures and helpers there, with executable
agent-tests/run-tests.sh. Installed environments, builds and caches
belong outside the export; the runner must reproduce required setup.
\end{quote}
\promptdivider

\begin{quote}\small\ttfamily\raggedright
The host independently executes and reviews the candidate; your
declaration cannot bypass verification. Use only public task evidence
and task-provided source; hidden benchmark files, reference solutions
and other candidates are unavailable.
\end{quote}
\promptdivider

\begin{quote}\small\ttfamily\raggedright
Choose the next action from the missing evidence and remaining budget.
Skills provide optional methods, not additional acceptance requirements.
Batch independent reads, follow artifact references and keep concise
evidence in update\_notes. Never emit submit\_candidate in the same
response as writes or exec: write the source/replay first, execute the
exact runner in a later response, then submit alone.
\end{quote}
\end{promptcard}

With the shared \texttt{execution-positive} ablation fallback,
\texttt{candidateWorkflow} replaces the optional repair sentence with:

\begin{promptcard}{02}{Publication fallback}
\begin{quote}\small\ttfamily\raggedright
The shared publication fallback requires a captured source repair that
passes your tests in a fresh sandbox, together with a reproducible
baseline failure. Preserve and submit incomplete artifacts honestly if
you cannot establish both controls.
\end{quote}
\end{promptcard}

\parabf{Branch strategies.}
At $K=3$, the three synthesis branches use the following exploration
strategies.
Each strategy changes exploration focus while retaining the original
public acceptance contract.

\begin{promptcard}{03}{Branch strategies}
\promptsubheading{Branch 1}
\begin{quote}\small\ttfamily\raggedright
Write a small public-behavior regression early, then develop a minimal
source repair and expand the suite with an independent small-input
oracle where possible.
\end{quote}
\promptdivider

\promptsubheading{Branch 2}
\begin{quote}\small\ttfamily\raggedright
Independently solve the complete task. Derive a small oracle or boundary
table from public examples, mathematics or documented behavior before
implementing the repair. Check one ordinary supported case as well as
the regression. Choose the repair independently of the test oracle.
Inspect uncertain expected values against the original contract; do not
inherit another solver's plan or assumptions.
\end{quote}
\promptdivider

\promptsubheading{Branch 3}
\begin{quote}\small\ttfamily\raggedright
Challenge likely incomplete implementations. Cover explicit requirements
overlooked by narrow tests and avoid adding rules the task does not state.
\end{quote}
\end{promptcard}

\parabf{Direction planning.}
The direction-planning prompt guides the first solver; peers retain
independent strategies. The prompt includes the following instructions.

\begin{promptcard}{04}{Direction planning}
\begin{quote}\small\ttfamily\raggedright
Plan a complete regression-and-repair workflow from the original public
contract and observed environment. Return one branch for the first
solver: write and run public-contract tests, implement a minimal source
repair as an internal positive control, rerun, and submit. Completion
requires baseline failure and repaired-source success; only tests are
exported.
\end{quote}
\promptdivider

\begin{quote}\small\ttfamily\raggedright
Independent peers derive their own approach without this plan. Cite
each requirement by taskLines in the numbered original task; the host
extracts quotes. Do not invent expected values, implementations, tooling
availability, or new acceptance rules. Prefer one focused inspection
that resolves a concrete unknown over a broad repository survey. The
host owns budgets and independent validation. Replanning must address
the observed invalid premise, preserving unaffected work.
\end{quote}
\end{promptcard}

\FloatBarrier
\subsection{Peer Cross Validation}
\label{app:testhelix-pcv}

\parabf{Evidence boundary.}
Generation uses the original public task and public source. Internal repairs
are hypotheses used as execution controls. Review runs in a fresh copy of
the original source with the generated tests; no generated source repair is
installed there. The host supplies bounded execution symptoms and checks
review citations against public evidence. Only tests are exported for
benchmark scoring. Reference Acceptance Rate during training provides host
feedback about the supplied reference; it does not expose the reference contents or the hidden
evaluation panel to the test generator. This objective remains distinct
from the final Joint Success Function metric.

\parabf{Independent semantic review.}
The reviewer receives the following instructions.
The final structured response supplies findings,
covered requirements, and execution checks; host validation derives
acceptance from verified blocking findings.

\begin{promptcard}{05}{Independent semantic review}
\begin{quote}\small\ttfamily\raggedright
You review independent regression tests against the original public
task. Your workspace contains the unchanged public baseline plus
generated tests. You cannot see generated source repairs or any hidden
oracle. Inspect the tests and relevant public interfaces. Decide
whether every assertion follows from explicit requirements or clearly
established repository behavior.
\end{quote}
\promptdivider

\begin{quote}\small\ttfamily\raggedright
Block only evidenced setup defects, unsupported assertions or missing
required behavior. Optional cases and speculative concerns are
suggestions; host derives acceptance from blocking findings. An excerpt
proves its existence, not your interpretation.
\end{quote}
\promptdivider

\begin{quote}\small\ttfamily\raggedright
For a blocking unsupported-assertion finding, supply falseRejection
describing a concrete allowed behavior, why it remains allowed given
established compatibility, whether the assertion is a justified control,
and publicBasis with taskLines or a repository line citation. Omission
from the task is not enough. If that claim cannot be established,
express the uncertainty as a suggestion; do not invent a counterexample
or automatically approve a failing suite.
\end{quote}
\end{promptcard}

\FloatBarrier
\subsection{Recursive Self Improvement}
\label{app:testhelix-development}

\subsubsection{Strategy Schema and Fixed Boundaries}

Table~\ref{tab:testhelix-policy-schema} documents the defaults and
schema bounds of the 11 strategy fields. Experiment configurations can
override the defaults. The experimental pair count settings are specified
separately in Appendix~\ref{sec:testhelix-experiment-details}; the RSI search ranges appear
in Appendix~\ref{app:testhelix-development}.

\begin{table}[htbp]
\centering
\caption{TestHelix strategy fields, defaults, and schema bounds.}
\label{tab:testhelix-policy-schema}
\small
\setlength{\tabcolsep}{4pt}
\begin{tabularx}{\linewidth}{@{}l l X@{}}
\toprule
Field & Schema default & Schema bounds \\
\midrule
\texttt{guidance} & Empty string & Reusable prose, at most 2,400 characters \\
\texttt{lessons} & Empty list & At most 20 lessons; at most four per explicit component \\
\texttt{skillRoutes} & Empty list & At most one enabled skill or null per context, seven contexts \\
\texttt{initialCandidates} & 2 & Integer from 1 to 5, no greater than the maximum \\
\texttt{maxCandidates} & 4 & Integer from 1 to 5 \\
\texttt{maxRounds} & 2 & Integer from 1 to 3 \\
\texttt{agentSteps} & 12 & Integer from 2 to 20 \\
\texttt{repairSteps} & 5 & Integer from 2 to 10 \\
\texttt{contextTokens} & 18,000 & Integer from 2,000 to 64,000 \\
\texttt{toolOutputChars} & 10,000 & Integer from 512 to 20,000 \\
\texttt{reviewMaxOutputTokens} & 2,048 & Integer from 128 to 16,384 \\
\bottomrule
\end{tabularx}
\end{table}

\parabf{Controlled optimization space.}
The fixed $K=3$ comparison sets both candidate fields to three and excludes
them from mutation. The remaining fields allocate the existing inference
budget or change reusable guidance; the execution model and its weights,
tool implementations, base skill library, execution validator, semantic
review protocol, and total task budget stay fixed. Configuration validation
imposes additional constraints: differential
challenges require at least two rounds, cross execution, semantic review,
and corrective feedback. A single proposal changes exactly one declared strategy
field. A lesson proposal changes exactly one item while preserving all
others. The host validates the schema and rejects embedded task identifiers
and code blocks in learned prose; the reflection instructions additionally
prohibit copying solutions or hidden patches.
The optimizer therefore influences execution through bounded strategy
decisions, while model interfaces and validation remain fixed. The effect
of coupling between the optimizer and execution model on learned guidance
is outside this study's scope.

\parabf{Lessons and routing.}
A lesson contains an identifier, one to eight literal retrieval triggers,
text of at most 1,000 characters, and optional component and capability
tags. Explicit components are planning, preparation, generation, repair,
and review. An update renders the text as applicability, action, and
observable verification. Retrieval uses the public task, current phase,
and host execution checkpoint; it selects at most two applicable lessons
within 1,600 characters for an agent turn, or 1,200 for planning.
The seven routing contexts are general, algorithm, framework, stateful,
performance, CLI, and visual. Routes select from the fixed enabled skill
library, and reviewer skills cannot be routed to candidate generation.
An empty route list leaves specialist loading to the agent; it does not
install a learned routing strategy.

\FloatBarrier
\subsubsection{Reusable Guidance and Strategy Reflection}

\parabf{Reusable strategy injection.}
The candidate receives the following user message only when
\texttt{guidance} is nonempty. Braces below identify runtime values.

\begin{promptcard}{06}{Reusable guidance}
\begin{quote}\small\ttfamily\raggedright
Reusable development-set guidance; apply only where the original task
supports it:\\
\{guidance\}
\end{quote}
\end{promptcard}

Retrieved lessons use the corresponding phase message and structured
lesson text.

\begin{promptcard}{07}{Retrieved lessons}
\begin{quote}\small\ttfamily\raggedright
Current \{component\} methods; original requirements take priority:\\
When: \{when\}\\
Action: \{action\}\\
Verify: \{verify\}
\end{quote}
\end{promptcard}

\parabf{Strategy reflection.}
The bounded reflection template supplies the strategy adapter with the
following instructions for revising reusable guidance from execution evidence.

\begin{promptcard}{08}{Strategy reflection}
\begin{quote}\small\ttfamily\raggedright
Improve a test-generation policy from development-set execution evidence.
Change exactly one policy component. For lessons, add, revise or remove
ONE item and preserve every other item. Write applicability, action and
an observable verification condition, with specific retrieval triggers.
Triggers match literal normalized phrases in public task instructions,
the current phase/reason, and workspace evidence: runnerPresent,
execution command and status, source repair and task coverage.
\end{quote}
\promptdivider

\begin{quote}\small\ttfamily\raggedright
Abstract failure mechanisms across tasks; do not copy names, paths,
literal expected values, solutions, task IDs, code, hidden patches, or
instructions to weaken validation. Preserve the public task as
authoritative. Do not alter model, tools, validators, or global budgets.
Explain the observed failure that motivates the mutation. Aggregate
training evidence is noisy; it is not proof of causality.
\end{quote}
\end{promptcard}

\parabf{Optimization and evaluation.}
AutoSaddler uses Claude Opus~4.6 to revise strategies and execute TestHelix
during optimization. We freeze the selected strategy's prompts and
configured fields, then use this same configuration with GPT~5.6 sol and
Kimi K3 for benchmark evaluation without further adaptation.

\subsubsection{Objective, Data Roles, and Search Configuration}
\label{app:rsi-config}

\parabf{Objective and isolation.}
For a task $i$, let $F_i(\pi)=1$ when the final test suite submitted by
strategy $\pi$ fails on the original program and passes on its supplied
reference repair; otherwise let $F_i(\pi)=0$. The objective is
$\frac{1}{|D|}\sum_{i\in D}F_i(\pi)$, the Reference Acceptance Rate
under the evaluation protocol. A missing suite, preparation error,
timeout, or unavailable required outcome receives zero in the fixed task
denominator. Training does not optimize Joint Success Function or inspect the hidden candidate
panel. The reference is accessible only to the host scorer, while the
generator receives public task information and bounded execution feedback.
The internal three branches produce one final submitted suite: $K=3$ is
not three independent scoring attempts or a pass@3 objective.

\parabf{Data assignment.}
The 600 optimization tasks, supplied in the supporting materials, are
partitioned into 500 proposal (training) tasks and 100 search development
(validation) tasks. All task assignments use seed 42. The proposal tasks provide execution
traces for diagnosis and strategy revision. Reference Acceptance Rate on
the search development set ranks candidates and can inform subsequent search;
it is consequently an optimization signal, not independent validation.
Another 100 tasks from the remaining 1,100 in the 1,700 task training pool
form a held-out optimization test set for release validation. These
three sets and the benchmark test set are disjoint. The assignment changes
neither the task pool nor the benchmark denominator.

\parabf{AutoSaddler adaptation.}
The AutoSaddler search uses Claude Opus~4.6 for candidate selection,
diagnosis and revision, and subsequent reflection through three agent
sessions. Opus~4.6 also executes TestHelix on the proposal and development
tasks, yielding one strategy search with a fixed execution model.
Its evolution DAG records parent relationships, edits, execution outcomes,
and reusable lessons. We adapt AutoSaddler~\citep{park2026autosaddler} to test
generation by invoking the test generation runner in single-reference evaluation
mode and mapping each completed task to the binary score above. The adaptation constrains
proposals to the strategy schema in
Appendix~\ref{app:testhelix-policy}, checks the changed field, and rejects
changes to the evaluator, tools, task inputs, or inference caps.

\parabf{Optimizer configuration.}
The corresponding AutoSaddler configuration uses
\texttt{train\_\allowbreak{}minibatch\_\allowbreak{}size=25}, \texttt{train\_\allowbreak{}eval\_\allowbreak{}attempts=1},
\texttt{train\_\allowbreak{}epochs=1}, \texttt{completed\_\allowbreak{}train\_\allowbreak{}iterations=20}, and
\texttt{max\_\allowbreak{}candidate\_\allowbreak{}proposals=40}. Both
\texttt{capability\_\allowbreak{}phase\_\allowbreak{}iterations} and
\texttt{capability\_\allowbreak{}phase\_\allowbreak{}epochs} are zero;
\texttt{skip\_\allowbreak{}session0}, \texttt{monotonic\_\allowbreak{}full\_\allowbreak{}train\_\allowbreak{}gate},
the adapter's \texttt{monotonic\_\allowbreak{}train\_\allowbreak{}gate}, and
\texttt{use\_\allowbreak{}merge} are false. Selection, mutation, and reflection have
separate session timeouts. Optimizer calls, tokens, failures, and elapsed
time are separate resource quantities from the generator's inference cap.

\begin{table}[htbp]
\centering
\caption{AutoSaddler settings for the TestHelix search.}
\label{tab:rsi-search}
\small
\setlength{\tabcolsep}{4pt}
\begin{tabularx}{\linewidth}{@{}l X@{}}
\toprule
Setting & Value and interpretation \\
\midrule
Training assignment & 500 proposal tasks and 100 search development tasks \\
Completion schedule & 20 completed rounds, 25 distinct proposal tasks per round, one epoch \\
Evaluation attempts & One final suite per task and strategy; seed 42 \\
Attempt ceiling & 40 attempts to complete 20 rounds; retries recorded separately \\
Optimizer and search execution model & Claude Opus 4.6 for both roles \\
Session timeouts & Selection 300 s; diagnosis and revision 1,800 s; reflection 600 s \\
Candidate health check & 60 s for strategy parsing, schema checks, and runner loading \\
Phase and selection & Capability iterations and epochs both zero; steering only; session 0 enabled \\
Full training gates & Both proposer and engine monotonic training gates disabled \\
Merge and release & Merge disabled; release validation restricted to one frozen challenger \\
\bottomrule
\end{tabularx}
\end{table}

\subsubsection{Initial Strategy, Mutable Fields, and Inference Limits}

\parabf{Initial strategy.}
The initial strategy eagerly requests three candidates and keeps
both candidate count fields fixed at three throughout optimization.
It uses two rounds, 12 agent steps, five repair steps, an 18,000 token
context limit, a 10,000 character tool output limit, and a 2,048 token
review output limit. Cross execution, semantic review, and corrective feedback
are enabled, with one differential challenge step and an execution positive
fallback. The 12 curated skills remain fixed; guidance, lessons, and skill routes initially are
empty. These settings instantiate the initial strategy.

\begin{table}[htbp]
\centering
\caption{Mutable strategy ranges and fixed inference caps.}
\label{tab:rsi-policy-budget}
\small
\setlength{\tabcolsep}{4pt}
\begin{tabularx}{\linewidth}{@{}l X@{}}
\toprule
Field or resource & Values \\
\midrule
\texttt{maxRounds} & $\{2,3\}$ \\
\texttt{agentSteps}; \texttt{repairSteps} & $\{10,12,14,16\}$; $\{3,5,7\}$ \\
\texttt{contextTokens} & $\{16{,}000,18{,}000,24{,}000,32{,}000\}$ \\
\texttt{toolOutputChars} & $\{8{,}000,10{,}000,14{,}000\}$ \\
\texttt{reviewMaxOutputTokens} & $\{2{,}048,4{,}096\}$ \\
Textual strategy fields & Guidance, lessons, and skill routes within the existing schema \\
Task profile caps & 48 model calls; 500,000 input and 100,000 output tokens;
900 s, further limited by public task deadlines \\
Model call concurrency & Two shared concurrent calls \\
Per call output limits & Generation 8,192 tokens; review initially 2,048 tokens \\
Execution and timeouts & Two execution repeats; 120 s model timeout; 180 s test timeout \\
Transport retry allowance & One retry, included in the shared budget and usage ledger \\
\bottomrule
\end{tabularx}
\end{table}

Each proposal changes one of the nine mutable fields; a lesson update
changes one lesson while preserving the rest. The six scalar ranges in
Table~\ref{tab:rsi-policy-budget} define the scalar search space.
Guidance, lessons, and routes obey the schema bounds and evidence
restrictions documented in Appendix~\ref{app:testhelix-policy}. The lower
bound of two rounds preserves
the differential challenge configuration. Increasing a component limit
does not increase the shared task cap: all generation, review, repair, and
feedback calls consume the same allowance. Provider reported usage is used
when available; otherwise accounting retains conservative reservations.
The task's public runtime limit also applies. For an official agent limit
$T$, the bridge reserves $\min(120,T/4)$ seconds for cleanup and another
20 seconds before passing a runtime cap to the Node process. The profile
cannot override that cap: the runner's default $T=600$ seconds yields a
460 second Node limit despite the 900 second profile setting. The test
execution limit is likewise the smaller of 180 seconds and the task's
public test timeout. All compared strategies retain the same task deadlines.
Execution repeats verify a
suite's behavior and do not grant extra independent test generation
attempts. The fixed prompts appear in Appendix~\ref{app:testhelix-policy}.

\subsubsection{Screening, Development Tracking, and Independent Release Validation}
\label{app:rsi-release}

Each completed round evaluates a parent on 25 previously unused proposal
tasks, produces a revision, and evaluates the child afresh on those same
tasks. The screen requires a strict increase in successful initial-to-reference
transitions and zero regressions among previously successful tasks. Only a child satisfying
both conditions receives evaluation on the 100 search development tasks.
The initial strategy receives the same development evaluation once.
AutoSaddler selects the strategy with the highest Reference Acceptance Rate
on the development set under Opus~4.6 and freezes it before release validation.
After this search, the RSI configuration in Table~\ref{tab:testhelix-rsi} uses the same frozen
strategy with GPT~5.6 sol and Kimi K3. Neither benchmark model supplies
feedback for strategy optimization, and the strategy is not adapted to
either model during benchmark evaluation. A development score need not
increase for a candidate to be evaluated or retained in the search history.

The release validation protocol compares exactly one frozen challenger with
the initial strategy on the separate 100 release validation tasks, using
Opus~4.6 to execute both strategies. The
promotion rule requires a positive paired difference in Reference Acceptance
Rate and a one sided exact McNemar $p\leq0.05$, with the direction fixed before evaluation.
For $w$ challenger only successes and $\ell$ initial only successes, the
test uses the upper binomial tail at $w$ conditional on $w+\ell$ discordant
pairs, with probability $1/2$ under the null. With no discordant pairs,
there is no evidence for promotion. The protocol reports both discordant
counts, both success counts, and 95\% and 99\% paired bootstrap intervals from
10,000 task resamples using seed 42. Each resample keeps the two outcomes for a task
together.

This modest validation set can lack power for small improvements.
Failure to pass the gate retains the initial strategy;
the validation results must not trigger further adaptive search on those
same tasks. The gate is separate from the benchmark scores.

\Needspace{12\baselineskip}
\parabf{Independent validation results.}
Table~\ref{tab:rsi-release} reports the paired comparison on the 100
release validation tasks. Reference Acceptance Rate increases from 48\% to
58\%, a gain of 10 percentage points. Both strategies succeed on 46 tasks and fail on
40; the remaining 14 tasks comprise 12 new successes and two regressions.
The positive gain and one sided exact McNemar $p=0.006470$ satisfy the
release criterion for the frozen strategy.

\begin{table}[H]
\centering
\caption{Independent release validation with Opus~4.6. Success counts use
100 paired tasks; confidence intervals describe the Reference Acceptance Rate gain in percentage
points (pp).}
\label{tab:rsi-release}
\small
\setlength{\tabcolsep}{5pt}
\begin{tabular}{@{}cccccc@{}}
\toprule
Initial & After RSI & \shortstack{New successes /\\regressions} &
\shortstack{One sided\\exact McNemar $p$} & 95\% CI (pp) & 99\% CI (pp) \\
\midrule
48/100 & 58/100 & 12 / 2 & 0.006470 & $[+3,+17]$ & $[+1,+20]$ \\
\bottomrule
\end{tabular}
\end{table}

\FloatBarrier

\section{Experimental Results and Analysis}
\label{sec:experimental-details}

The following sections follow the research questions in the main text:
baseline performance, patch diversity, test failure mechanisms, and
TestHelix component comparisons. Metric definitions appear in
Appendix~\ref{sec:evaluation-details}; TestHelix strategies and optimization
procedures appear in Appendix~\ref{app:testhelix-policy}.

\subsection{RQ1: Benchmarking Coding Agents}
\label{app:rq1-details}

\subsubsection{Baseline Configurations}
\label{sec:baseline-configurations}

Claude, Kimi, and Qwen models use Claude Code, Kimi Code, and Qwen Code,
respectively. GPT 5.6 sol uses Codex, DeepSeek models use DeepSeek Harness,
and GLM models use ZCode. Table~\ref{tab:benchmark300-main} retains each model
family's configuration with the highest Joint Success Function.
Table~\ref{tab:additional-baseline}
reports the remaining eight configurations, including GPT 5.4 with Terminus-2,
sorted by descending Joint Success Function. Earlier versions and model variants support
comparisons within families; GPT 5.4 with Terminus-2 provides
an additional harness baseline. Baseline comparisons and correlations include all
fourteen configurations in Tables~\ref{tab:benchmark300-main}
and~\ref{tab:additional-baseline}.

Native baselines use the latest harness releases available by August 15, 2026,
except for ZCode as specified below. They have no experiment-imposed inference
budget cap and retain the public task execution limits.
They do not share the TestHelix inference caps;
budget matching applies to the five TestHelix configurations in RQ4.
The cutoff releases are Claude Code~\href{https://github.com/anthropics/claude-code/releases/tag/v2.1.233}{2.1.233},
Kimi Code CLI~\href{https://github.com/MoonshotAI/kimi-code/releases/tag/\%40moonshot-ai/kimi-code\%400.36.1}{0.36.1},
Qwen Code~\href{https://github.com/QwenLM/qwen-code/releases/tag/v0.21.12}{0.21.12},
Codex CLI~\href{https://github.com/openai/codex/releases/tag/rust-v0.147.0}{0.147.0},
DeepSeek Harness~\href{https://www.npmjs.com/package/@deepseek-ai/dsh/v/0.1.0-rc.6}{0.1.0-rc.6},
and Terminus-2 bundled with Harbor~\href{https://github.com/harbor-framework/harbor/releases/tag/v0.21.0}{0.21.0}.
These are the latest stable releases by the cutoff, except for DeepSeek Harness,
whose available release is a release candidate.
ZCode uses the August 28, 2026 release.
Except for the explicitly labeled \texttt{-low} configuration, we use the
highest supported reasoning setting: \texttt{max} for Claude, Kimi K3,
GPT 5.6 sol, DeepSeek, and GLM;
\texttt{xhigh} for Qwen3.8 Max and GPT 5.4. Kimi K2.7 Code uses mandatory
thinking, and Qwen3.7 Plus uses thinking mode without an additional
thinking-token cap.

\begin{table}[!htbp]
\centering
\caption{\textbf{Additional coding agent configurations on TestPrism.}}
\label{tab:additional-baseline}
\begingroup
\footnotesize
\setlength{\tabcolsep}{3pt}
\begin{tabular*}{\linewidth}{@{\extracolsep{\fill}}lrrrrrrr}
\toprule[1pt]
\textbf{Model}
& \multicolumn{1}{c}{\textbf{Reference}}
& \multicolumn{3}{c}{\textbf{Candidates}}
& \multicolumn{2}{c}{\textbf{Coverage}}
& \multicolumn{1}{c}{\textbf{Joint}} \\
\cmidrule(lr){2-2}\cmidrule(lr){3-5}\cmidrule(lr){6-7}\cmidrule(l){8-8}
& RAR & VAR & IRR & $H$ & $\Delta\mathcal{C}$ & $\mathrm{EC}$ & \textbf{JSF} \\
\midrule
Kimi K2.7 Code & 51.67\% & 75.33\% & 55.33\% & 63.80\% & 78.33\% & 49.67\% & \cellcolor{tableaccent}21.67\% \\
Qwen3.7 Plus & 48.67\% & 81.00\% & 48.67\% & 60.80\% & 72.67\% & 47.67\% & \cellcolor{tableaccent}19.67\% \\
Claude Opus 5 & 55.00\% & 77.67\% & 74.33\% & 75.96\% & 71.67\% & 47.33\% & \cellcolor{tableaccent}19.00\% \\
GPT 5.6 sol -low & 35.33\% & 50.33\% & 81.67\% & 62.28\% & 58.33\% & 45.67\% & \cellcolor{tableaccent}14.67\% \\
DeepSeek V4.1 Flash & 39.67\% & 49.67\% & 78.00\% & 60.69\% & 65.33\% & 44.67\% & \cellcolor{tableaccent}13.00\% \\
DeepSeek V4 Flash & 37.00\% & 43.00\% & 83.67\% & 56.81\% & 61.33\% & 45.33\% & \cellcolor{tableaccent}11.67\% \\
GLM 5.2 & 39.33\% & 44.33\% & 90.33\% & 59.47\% & 60.67\% & 42.67\% & \cellcolor{tableaccent}10.00\% \\
GPT 5.4 & 23.67\% & 39.33\% & 74.67\% & 51.52\% & 56.67\% & 39.33\% & \cellcolor{tableaccent}8.33\% \\
\bottomrule[1pt]
\end{tabular*}
\endgroup\end{table}

\subsubsection{Metric Correlations}
\label{sec:coverage-analysis-details}

RQ1 computes Pearson correlations across the aggregate scores in
Tables~\ref{tab:benchmark300-main} and~\ref{tab:additional-baseline}.
Each of the fourteen coding agent
configurations contributes one paired observation with equal weight.
We compare $\Delta\mathcal{C}$ with Entry Coverage and relate each to
Reference Acceptance Rate and Joint Success Function. Coverage targets and collection conventions follow
Appendix~\ref{sec:coverage-details}.
We compute $H$ from Valid Acceptance Rate and Invalid Rejection Rate at the
precision shown in the tables; its Pearson correlation with Joint Success Function is $r=0.83$.

Spearman correlations show the same positive associations: $\rho=0.88$
between the coverage measures, $0.93$ and $0.83$ with Reference Acceptance Rate, and $0.90$ and
$0.96$ with Joint Success Function, for $\Delta\mathcal{C}$ and Entry Coverage, respectively.
These statistics describe variation across configurations evaluated on the
same task set; the unit of analysis is a configuration aggregate.

\FloatBarrier

\subsection{RQ2: Patch Diversity and Joint Success}
\label{sec:patch-count-diagnostic}

Figure~\ref{fig:rq2-patch-count} shows Joint Success Function and candidate diversity.
The score equally weights the six configurations in
Table~\ref{tab:benchmark300-main} on the same 50 tasks randomly sampled
from the 300 benchmark tasks using seed 42.
We use the same candidate generation and verifier validation procedure to
expand each task's standard ten implementation panel to 15 implementations:
the reference, seven additional valid alternatives, and seven invalid
candidates. These additions leave the panels used in the baseline and
TestHelix comparisons unchanged. The submitted tests and the 50 task subset
remain fixed while candidate sets grow under three addition schedules.
The valid and invalid sets in the joint criterion are the selected diagnostic
candidate sets, with the same labeling and verdict requirements.

\paragraph{Diversity convention.}
We use the AST-feature Jaccard distance $d_{\mathrm{AST}}$ defined in
Appendix~\ref{app:diversity-conversion}.
For a diagnostic panel $S$ with at least two candidates, diversity is the arithmetic
mean of $d_{\mathrm{AST}}(p,q)$ over its distinct unordered pairs; a singleton
is assigned zero. For each task, we add candidates uniformly at random without
replacement from the label pool required by the addition schedule, using seed 42.
The resulting order defines nested candidate prefixes, each retaining the reference.
Figure~\ref{fig:rq2-patch-count}(a) computes diversity on these realized prefixes
and averages equally over the 50 tasks. This descriptive mean includes pairs
across labels. The group admission
criteria apply to standard ten-implementation panels; we do not reapply
them to intermediate diagnostic subsets.
Panel (c) re-expresses the values in panels (a) and (b), pairing them by
addition schedule and patch count $k$. It is a descriptive visualization
parameterized by $k$, rather than an estimate of the relationship between
diversity and joint success at a fixed panel size.

\paragraph{Score calculation.}
For configuration $m$ and task $i$, let $c_{im}(p)$ indicate a correctly
recorded verdict: acceptance for a valid candidate or rejection for an invalid
one. Missing outcomes and recognized errors give $c_{im}(p)=0$.
Beyond the valid reference $g_i$, let $n_i^V$ and
$n_i^I$ count the available valid and invalid candidates for task $i$.
Let $s_{im}^V$ and $s_{im}^I$ count configuration $m$'s correct verdicts in
the same candidate pool, and let $F_{im}$ indicate its successful
initial to reference transition on task $i$.
Under independent uniform selection without replacement of $a$ additional
valid candidates and $b$ invalid candidates, let $\mathcal{P}_{a,b}$ contain
these candidates and the reference for each task. Averaging uniformly over
the six configurations and 50 tasks, the theoretical expected joint score is
\[
\overline{\text{Joint Success Function}}(a,b)=\frac{1}{6\cdot50}\sum_{m=1}^{6}\sum_{i=1}^{50}
 F_{im}\frac{C(s_{im}^V,a)}{C(n_i^V,a)}\frac{C(s_{im}^I,b)}{C(n_i^I,b)},
\]
where $C(n,j)$ counts subsets of size $j$ from $n$ elements, with
$C(n,j)=0$ when $j>n$. The total patch count is $k=1+a+b$;
the initial program is always checked and is not counted as a patch.
The calculation requires $a\leq n_i^V$ and $b\leq n_i^I$ for every task.
No task is removed as $k$ increases. Each of the six configurations receives
equal weight. For these expanded diagnostic pools,
$n_i^V=n_i^I=7$.
This expression averages over all permitted subsets. A uniformly sampled
addition order yields a realized sequence of prefixes; measurements on a
finite collection of such prefixes can differ from the exact subset expectation.
The reported score checkpoints use five valid additions
($k=6$), five invalid additions ($k=6$), and nine alternating additions
($k=10$).

\paragraph{Three addition schedules.}
Write $k=1+j$, where $j$ counts added candidates.
The valid addition curve uses $(a,b)=(j,0)$;
the invalid addition curve uses $(a,b)=(0,j)$, both for $0\leq j\leq7$.
Both retain the valid reference. The alternating curve follows the reference
with an invalid candidate, then a valid candidate, continuing to alternate:
$(a,b)=(\lfloor j/2\rfloor,\lceil j/2\rceil)$ for $0\leq j\leq14$.
Including the reference, its label sequence is valid, invalid, valid, invalid,
and so on. At $k=15$, the alternating schedule contains eight valid and seven
invalid implementations. Within each addition schedule, the curve evaluates
Joint Success Function at patch count $k$ under its corresponding schedule, without repeating
a candidate.

At $k=1$, the empty product is one and the metric equals the six configurations'
mean Reference Acceptance Rate, denoted $\overline{\text{Reference Acceptance Rate}}$. Each additional
required verdict can only preserve or remove success, so expected Joint Success Function is
nonincreasing along each schedule. Within the expanded fifteen implementation
panels, for $0\leq a\leq7$ and $0\leq b\leq7$,
$\overline{\text{Reference Acceptance Rate}}=\overline{\text{Joint Success Function}}(0,0)\geq\overline{\text{Joint Success Function}}(a,b)\geq\overline{\text{Joint Success Function}}(7,7)$.
Mean Reference Acceptance Rate is therefore the upper bound under the weakest criterion, and the full
diagnostic panel gives the lower bound among its subsets. This ordering does not
establish a universal bound on test quality or isolate the effect of candidate
diversity selection.

Mean Joint Success Function starts at 49.33\% and reaches 38.10\% after five valid
additions, 30.49\% after five invalid additions, and 24.33\% after nine
alternating additions. The complete pool endpoints are 37.33\%, 29.33\%,
and 23.33\% for the valid, invalid, and alternating schedules.
Complete pools include all admitted candidates; comparisons at different
panel sizes impose distinct verdict requirements.

All 50 tasks remain in each configuration's denominator. Missing or
erroneous required outcomes receive no success credit, following the main
evaluation protocol. Consequently, a decrease can reflect either a candidate
disagreement or an additional missing or erroneous required observation.

\paragraph{Marginal acceptance and rejection.}
Across the fourteen baselines in Tables~\ref{tab:benchmark300-main}
and~\ref{tab:additional-baseline}, Reference Acceptance Rate exceeds
Joint Success Function by 15.33 to 36.00 percentage points: 53.07\% to 74.58\% of reference
successes fail the joint criterion.
In Table~\ref{tab:benchmark300-main}, Claude Fable 5.1 leads both
Reference Acceptance Rate and Joint Success Function,
yet achieves joint success on only 84 of its 179 reference successes.
Qwen3.7 Plus has the highest Valid Acceptance Rate at
81.00\% but the lowest Invalid Rejection Rate at 48.67\%; GLM 5.2 has the
highest Invalid Rejection Rate at 90.33\% but a Valid Acceptance Rate of
44.33\%. Their Joint Success Function values are only 19.67\% and 10.00\%,
respectively. Strong acceptance or rejection alone therefore does not
establish reliable discrimination across the panel. These aggregate rates
describe different behavioral profiles; they do not identify the overlap of
valid and invalid candidate failures within reference successes. RQ1 relates
coverage to reference and joint success.

\FloatBarrier

\subsection{RQ3: Analysis of Test Generation Failures}
\label{sec:error-taxonomy-details}

\subsubsection{Annotation Units}

A failed trajectory has task-level joint success indicator $S_i=0$ under
the evaluation protocol. Using seed 42, we sample 50 such trajectories from the configuration
with the highest Joint Success Function in each of six model families
(Table~\ref{tab:benchmark300-main}), yielding 300 trajectories. Excluding
14 timeouts leaves 286 for expert annotation and
analysis. Each observation comprises a task, model configuration, submitted
test suite, and evaluation attempt. The reported proportions use these 286
trajectories as their denominator: 99 coverage gaps (34.6\%), 143 faulty
checks (50.0\%), and 44 execution logic errors (15.4\%).

The examples illustrate the mechanism definitions in
Figure~\ref{fig:error-taxonomy}(b) and the behavior of individual test submissions.
Disputed candidate labels do not become evidence of an erroneous generated
test solely because the generated verdict disagrees with trusted tests.

\subsubsection{Expert Annotation and Agreement}

Two experts independently annotate all 286 retained trajectories using the
seven error subtypes defined below. They inspect task requirements, submitted
test suites, initial and reference executions, candidate outcomes, and
supporting patch evidence. Cohen's $\kappa=0.81$ is computed from these
independent subtype labels before adjudication. For each disagreement, a third
expert joins both experts in discussion to determine the final label. We
aggregate the final labels into the three parent categories reported in
Section~\ref{sec:rq3-errors} and Figure~\ref{fig:error-taxonomy}(b).

\subsubsection{Subtype Boundaries}

\paragraph{Coverage Gaps.}
Missing Inputs concerns omitted inputs, states, configurations, or
combinations within an exercised operation. Untested Requirements concerns
entire required behaviors or entry points without acceptance checks.

\paragraph{Faulty Checks.}
Weak Checks and Wrong Criteria distinguish insufficient checks from
incorrect or unsupported behavioral criteria.
Overfitting enforces incidental implementation details beyond
the task contract.

\paragraph{Execution Logic Errors.}
Fixture \& API Errors concern errors in constructing test fixtures or using
APIs within the test process. Command \& Parsing Errors concern errors in
the test's command execution or output collection and parsing.
External installation, network, and container failures remain infrastructure
issues. Incorrect expectations about candidate behavior belong to Faulty
Checks.

\subsubsection{Connection to Candidate Panel Evaluation}

A direct single reference blind spot requires the same submitted suite to
fail on the initial program, pass on the reference, and accept a candidate
whose contract violation is supported by local evidence. Missing Inputs and
Weak Checks provide examples of this pattern: the same suite passes the
reference while accepting an invalid alternative.

Reference failures caused by execution logic errors or other mechanisms
cannot explain a decline conditional on reference success
without additional paired evidence. Trusted candidate labels guide witness
selection but do not establish correctness for every input. These observations
identify mechanisms consistent with the aggregate evaluation gaps; they do
not establish causal attribution for aggregate model scores or estimate the
prevalence of each mechanism across the benchmark.

\FloatBarrier
\Needspace{8\baselineskip}

\subsection{RQ4: Evaluating TestHelix}
\label{sec:testhelix-experiment-details}

\subsubsection{Experimental Controls}

Table~\ref{tab:testhelix-rsi} includes native agent comparators for each
model. The baseline comparison appears in Tables~\ref{tab:benchmark300-main}
and~\ref{tab:additional-baseline}. The following protocol defines five
TestHelix configurations, denoted A through E in table order. The first four
use the initial test generation strategy; the fifth uses a strategy improved
through Recursive Self Improvement (RSI), with three test and repair pairs.
Strategies remain frozen during benchmark evaluation. Here $K$ counts internal
test and repair pairs, independently of the hidden evaluation panel size.

\parabf{Single.}
Configuration A uses one test and repair pair ($K=1$), with initial state
checks, an own repair control, and self correction. Its suite qualifies locally when it
reproducibly fails on the initial state and passes on its own repair.
Peer Cross Validation is disabled.

\parabf{Heterogeneous Synthesis with random submission.}
Configuration B uses Heterogeneous Synthesis to develop three pairs
($K=3$) with distinct synthesis directions. Each pair is developed independently
using local execution checks for self correction. Cross execution, independent
semantic review, corrective feedback, and suite merging remain disabled.
After local correction, one suite is selected uniformly at random from the
locally qualified pairs, using seed 42 fixed before evaluation. This
configuration never enters peer cross validation, and selection uses no hidden
evaluation outcomes. A task with no qualified suite remains in the evaluation set and
receives no success credit; it is not replaced or resampled.

\parabf{Peer Cross Validation.}
Configuration C retains three pairs and combines the cross execution,
semantic review, and corrective feedback operations in Section~\ref{sec:testhelix}.
Peer repairs qualify as witnesses through local consistency checks, with
distinct lineage and source content preventing duplicate support.
Independent review interprets the task, original source, tests, and bounded
execution symptoms without seeing generated repairs. Reviewed peer
counterexamples guide correction. When distinct qualified repairs pass the
existing suites and both suites pass review, differential challenges search
for missing behavioral distinctions. Revised suites undergo fresh execution
and review. Selection uses witness support, and compatible suites with common
passing witnesses may be merged. The B to C comparison evaluates this
complete feedback process, including its selection and merging rules.

\parabf{Pair count.}
Configuration D uses five pairs ($K=5$) with the same peer cross validation process
and initial strategy as C. The C to D comparison varies pair count
under the same total inference limit. Five pairs appear only in this
comparison, without additional random submission or RSI configurations.

\parabf{RSI.}
Configuration E uses peer cross validation with three pairs and one improved,
frozen test generation strategy. AutoSaddler uses Opus~4.6 for both strategy
revision and task execution during optimization; GPT~5.6 sol and Kimi K3
then share the selected strategy for benchmark evaluation without further
adaptation. Both \texttt{initialCandidates} and \texttt{maxCandidates}
are fixed at three throughout strategy search, validation, and benchmark
evaluation. The remaining guidance and allocation fields can evolve within
the same total budget. The RSI increment compares E with C, which uses
the initial strategy at the same pair count. The frozen strategy passes
the independent Opus~4.6 release validation in
Appendix~\ref{app:rsi-release}. Benchmark tasks and their hidden panels do
not enter strategy updates.

\parabf{Shared controls and submission.}
Within each model group, all five configurations use the same base model, public
task access, evaluator, tools, base skill library, and total inference limits.
Strategy search uses training tasks and a disjoint validation gate; both task
sets exclude the final benchmark. Learned allocations remain subject to the
same total inference limits as the initial strategy. The budget is shared
across pairs rather than multiplied by $K$; all requested pairs begin in
the initial round. The run configuration comprises the planning settings,
execution repetitions, random seed, pair count, and strategy.
Disabling feedback excludes it from generation and selection, including
cached evidence and prompts. Configurations A and B require local consistency for
submission. Configurations C, D, and E also require semantic approval; if no
common peer witness is available, an approved suite with reproducible initial
failure and a passing own repair remains eligible. Internal qualification
does not provide success credit in the hidden benchmark evaluation.

\parabf{Reporting and interpretation.}
All \benchTaskCount{} tasks remain in the Reference Acceptance Rate and
Joint Success Function denominators.
Valid Acceptance Rate measures acceptance of the four valid alternatives,
excluding the reference. This rate and Invalid Rejection Rate use the
observed candidate denominator
within each task and average the defined task rates with equal weight.
Coverage uses observed, nonempty measurement denominators;
missing coverage is not zero. Within each model group, all five
configurations use the same benchmark task set.
These configurations share inference limits. Inference consumption includes
every role, retry, and correction call;
strategy development and benchmark inference are distinct accounting scopes.
Equal limits do not imply equal consumed compute.

The A to B comparison evaluates the complete Heterogeneous Synthesis design,
which expands both
the number of independently developed test and repair pairs and the range
of exploration directions. Guided by the branch prompts, the LLM determines
concrete task-specific directions, including complementary behaviors and
corner cases. Additional pairs provide space to pursue these directions
within the same total inference cap. This comparison therefore measures
the overall benefit of heterogeneous expansion relative to Single.
The B to C comparison evaluates the complete peer cross validation process,
including cross execution, semantic review, and corrective feedback.
The C to D comparison measures pair count under a fixed budget, and
the C to E comparison measures RSI at fixed $K=3$. The RSI increment is
therefore relative to the three pair peer cross validation configuration, not the adjacent
five pair row.

\clearpage
\section{Limitations}
\label{sec:limitations}

Reliable test generation spans a broad space of task specifications, inputs,
implementations, and execution environments. This study focuses on one part
of that problem: whether tests written by coding agents distinguish valid
and invalid implementations under fixed task contracts. TestPrism evaluates
ten labeled implementations per task, and Joint Success Function measures joint success within these
finite candidate patch sets. The comparisons concern complete deployed agent
configurations, while the TestHelix ablations examine component contributions
within each model group. The taxonomy complements these results by
characterizing recurring failure mechanisms in 286 analyzed failed trajectories,
after excluding 14 timeouts from 300 trajectories sampled equally from the
highest Joint Success Function configuration in each of six model families.
Failed trajectories have a per-task joint success indicator of zero. The category
proportions describe the retained sample and exclude the timeout trajectories.

\parabf{Source gold tests.}
We select established public benchmarks with documented test validation,
including the widely adopted SWE bench and Terminal Bench families
(Appendix~\ref{app:source-test-validation}). Their original quality controls
provide the foundation for task and verifier quality. Building on this
foundation, our second-stage validation checks adapted verifiers through
execution and independent expert review for requirement alignment and
potential reference overfitting; tasks with unresolved concerns are excluded.
Our scope is evaluating generated tests across implementations under these
validated task contracts. Independently re-establishing the correctness of
each upstream benchmark is outside this study.

\parabf{Panel composition and diversity.}
The prevalence of valid and invalid human patches is difficult to establish
and can vary across tasks and development settings. Our equal allocation
provides a balanced evaluation design without estimating this prevalence.
The diversity contributed by a candidate also depends on task semantics,
implementation choices, defect patterns, and its relationship to candidates
already in the panel. Consequently, a valid to invalid ratio alone does not
determine diversity, and equal counts do not guarantee equal difficulty for
acceptance and rejection.
Our admission checks use mean AST distance for the valid group and failed-check
sets with conditional AST filtering for the invalid group. They do not establish exhaustive
semantic coverage or an optimal candidate set. Jaccard distance summarizes
overlap in the chosen AST feature representation and can differ from semantic
distance; its mean over a panel need not increase with every added patch.
This work introduces Joint Success Function and candidate patch sets with explicit diversity
controls. Optimizing the label ratio, panel size, or candidate selection is
outside its scope.

Extending evaluation to evolving specifications, broader implementation
spaces, and additional execution environments remains a direction for future work.

\end{document}